\documentclass[12pt,a4paper]{article}

\usepackage[utf8]{inputenc}
\usepackage{fullpage}
\usepackage{hyperref}
\usepackage{amssymb}
\usepackage{amsthm}
\usepackage{graphicx}
\usepackage[outdir=./]{epstopdf}
\usepackage{color}
\usepackage{enumitem}
\usepackage{empheq}
\usepackage{cite}
\usepackage{slashed}
\usepackage{amsmath}
\usepackage{amsfonts}
\usepackage{cancel}
\usepackage{float}
\usepackage{physics}
\usepackage{graphicx}
\usepackage{caption}
\usepackage{subcaption}
\usepackage{tikz}
\usepackage[compat=1.1.0]{tikz-feynman}

\numberwithin{equation}{section}
\newcommand{\s}[1]{\slashed #1}
\newcommand{\LP}{\mathrm{LP}}
\newcommand{\NLP}{\mathrm{NLP}}
\newcommand{\savg}[1]{\overline{\left|#1\right|}^2}

\restylefloat{figure}

\usepackage{cleveref}
\crefname{section}{Section}{Sections}
\crefname{appendix}{Appendix}{Appendices}

\begin{document}

\titlepage

\begin{flushright}
    MS-TP-23-47\\
    TUM-HEP-1486-23
\end{flushright}

\vspace*{1.2cm}

\begin{center}
    {\Large \bf Soft-photon spectra and the LBK theorem}

\vspace*{1.5cm} \textsc {R. Balsach$^{ \flat, (a)}$, D.
Bonocore$^{ \sharp, (b)}$, A.
Kulesza$^{ \flat, (c)}$} \\

\vspace*{1cm}

$^\flat$ Institut f\"{u}r Theoretische Physik, Westf\"{a}lische
Wilhelms-Universit\"{a}t M\"{u}nster, Wilhelm-Klemm-Stra\ss e 9,
D-48149 M\"{u}nster, Germany

\vspace*{0.2cm}
$^\sharp$ {Technical University of Munich, TUM School of Natural Sciences, \\
Physics Department T31,
    James-Franck-Straße 1, D-85748,
    Garching, Germany}\\

\end{center}

\vspace*{7mm}

\begin{abstract}
    \noindent
    The study of  next-to-leading-power (NLP) corrections in soft
    emissions
    continues to attract interest both in QCD and
    in QED. Soft-photon spectra in particular provide a clean case-study for
    the experimental verification of the Low-Burnett-Kroll (LBK) theorem.
    In this paper we study the consistency of the LBK theorem in the context of
    an ambiguity arising from momentum-conservation constraints in the
    computation of non-radiative amplitudes. We clarify that this ambiguity
    leads to various possible formulations of the LBK theorem, which are all
    equivalent up to power-suppressed effects (i.e. beyond the formal accuracy
    of the
    LBK theorem). We also propose a new formulation of the LBK theorem with a
    modified shifted kinematics which facilitates the numerical
    computation of non-radiative amplitudes with publicly available tools.
    Furthermore, we present numerical
    results for soft-photon spectra in the associated production of a muon pair
    with a photon, both in $e^+e^-$ annihilation and proton-proton collisions.
\end{abstract}

\vspace*{\fill}
{\footnotesize
\noindent
(a) rbalsach@uni-muenster.de\\
(b) domenico.bonocore@tum.de\\
(c) anna.kulesza@uni-muenster.de
}


\newpage

\begingroup
\hypersetup{hidelinks}
\tableofcontents
\endgroup


\section{Introduction}
\label{sec:intro}

Perturbative calculations are the cornerstone of theoretical predictions for
high energy physics experiments. The
expansion in the
coupling constant is arguably the most important example,
with the expansion terms denoted as leading order (LO), next-to-leading order
(NLO), and so forth. For processes
involving
several scales, a larger number of dimensionless parameters can be small in
particular kinematic limits, hence other
expansions are possible. A case that has received substantial attention is the
study of power corrections to the strict soft and/or
collinear limit, whose expansion terms are conventionally denoted as
 leading power (LP), next-to-leading power (NLP), etc.

The theoretical foundations of NLP emissions date back to
the theorems
of Low,
Burnett and Kroll
(LBK) \cite{Low:1958sn, Burnett:1967km} (see also \cite{Gell-Mann:1954wra,
Bell:1969yw}), which
continue to be reformulated and generalized also in
the recent years\footnote{Soft theorems are an active field of
research also at LP, see e.g.
    \cite{Hannesdottir:2019opa, Agarwal:2021ais,
    McLoughlin:2022ljp,
        Feal:2022iyn, Chen:2023hmk,
        Herzog:2023sgb,
        Ma:2023gir}.}
\cite{DelDuca:1990gz, Laenen:2008gt, Bonocore:2020xuj, Cachazo:2014fwa,
Strominger:2017zoo,
Casali:2014xpa, Bern:2014oka,
Larkoski:2014bxa,
Luo:2014wea, He:2014bga, Bonocore:2015esa, Beneke:2019oqx, Gervais:2017yxv,
Laddha:2018myi,
Laenen:2020nrt,
DelDuca:2017twk,
Bonocore:2021cbv, Beneke:2021umj, Beneke:2022pue,
Engel:2021ccn, Engel:2023ifn, Engel:2023rxp, Czakon:2023tld,
Lebiedowicz:2021byo, Lebiedowicz:2023mlz, Lebiedowicz:2023ell}. In
particle phenomenology, these subleading
effects have mainly
attracted
attention due to their potential relevance for QCD
resummation\footnote{
    NLP effects are also relevant for the numerical stability of differential
    NNLO calculations, both in QCD \cite{Moult:2017jsg, Boughezal:2018mvf}  and
    in QED
    \cite{Banerjee:2021mty, Broggio:2022htr}.}. Indeed, it is well-known that
    infrared divergences due to unresolved soft
and collinear radiation yield logarithms in the cross-section that
 become large when approaching some kinematic threshold, thus spoiling
the predictive power of finite-order perturbation theory. The goal of the
traditional (i.e.
LP) resummation program is to reorganize the towers of these logarithms at a
given logarithmic accuracy
to all-orders in perturbation theory.
In this context, subleading  corrections due to emissions of gluons (and
quarks) give rise to NLP
logarithms which, although power-suppressed in the threshold limit, could give
significant
contribution to the cross-section. In the last decade, a considerable
effort has been invested in this direction \cite{Moult:2018jjd,
Bahjat-Abbas:2019fqa, Liu:2020tzd, Beneke:2018gvs,
Cieri:2019tfv,
Agarwal:2023fdk, Broggio:2023pbu, Ravindran:2022aqr, Sterman:2022lki}.

The soft limit in the photon bremsstrahlung \cite{Bethe:1934za, Bloch:1937pw,
    Landau:1953um, Jackson:1998nia, Feal:2018bru} provides another probe of NLP
effects.
 In this case, the study of the
photon
spectrum
 gives direct access to the individual terms of NLP soft theorems, unlike
the QCD resummation case, where one is blind to the energy of the undetected
gluon since
its momentum is integrated over the
whole phase space.
In fact, although the LBK theorem is very old and the conditions that
ensure
the soft limit for a given process are known in terms of a well-defined
hierarchy of scales, to the best of our knowledge, no study
in the
literature has
  studied
numerically  what is the resolution in energy and momentum of a soft photon
that one
has to  reach in order for NLP effects to be measurable.

This question is not a mere theoretical exercise. Soft-photon spectra in
hadronic decays have been puzzling physicists for years. The discrepancy
between the LP prediction and the observed yields of photons produced together
with hadrons is outstanding and   the results of the measurements
remain not understood at present \cite{DELPHI:2005yew, DELPHI:2007nmh,
DELPHI:2010cit,
Chliapnikov:1984ed, EHSNA22:1991sdp, SOPHIEWA83:1992czx, WA91:1997cnv,
Belogianni:2002ic, Belogianni:2002ib}. Moreover, there are plans for an upgrade
of the
ALICE
detector at the Large Hadron Collider that would enable the possibility of
scrutinizing photons at ultra-soft energies \cite{Adamova:2019vkf,
ALICE:2022wwr}. In order
to shed light on these discrepancies
and correctly interpret data from   future measurements, it is therefore of
the utmost importance to have reliable
theoretical predictions, including also NLP
corrections as first proposed in this context in
\cite{Bonocore:2021cbv}. With this long-term goal in mind, in this paper we
study the tree-level form of the LBK theorem for the production of a photon in
association with a $\mu^+\mu^-$ pair in $e^+e^-$ and $pp$ collisions.

To analyse the soft-photon spectrum, one has to  evaluate the expressions
 given by the LBK theorem.
In fact,
several issues must be  addressed, both analytically and
numerically.
Most notably, as it has been already pointed out in \cite{Bonocore:2021cbv},
the traditional form of the theorem expressed through derivatives of the
non-radiative amplitude is not optimal. Indeed, the non-radiative amplitude
depends on a set
of
unphysical momenta that violates momentum conservation when the soft-photon
momentum $k\neq 0$. This is
problematic since,  by definition, photon spectra are calculated for a
non-vanishing momentum $k$. To overcome this issue, the strategy proposed
first for two massless legs in
\cite{DelDuca:2017twk} \footnote{See also \cite{vanBeekveld:2019prq} and
the recent \cite{vanBeekveld:2023gio}.}   and then
generalized in
\cite{Bonocore:2021cbv} for an arbitrary number of (massless or massive) legs,
 appears promising.
The strategy relies on
removing the derivatives of the non-radiative amplitude by computing such
amplitude on
momenta which are slightly shifted in value.
Remarkably, the sum of these momenta
shifts is equal to the soft-photon momentum, so that momentum conservation is
restored. The price to pay for
this
trick is that the shifted momenta do not fulfill the on-shell conditions. This
issue
prevents  the
calculation of the non-radiative amplitude with most of the available public
tools which can be used for the numerical evaluation of matrix elements. It is
one of the goals
of this paper to explicitly show how the momenta of the external particle can
be
kept on-shell by
proposing a modified version of the shifts that are equivalent to the ones
discussed in \cite{Bonocore:2021cbv} up to NNLP corrections.

The  observation of the dependence of the amplitude on non-physical momenta at
NLP is an old one, and it was
first discussed by Burnett and Kroll \cite{Burnett:1967km}. More recent and
detailed discussions on this aspect can be found in \cite{Gervais:2017yxv} (see
also \cite{Engel:2021ccn}).
Despite the long history and the large body of papers  which studied,
reformulated and generalized the LBK theorem, the issue of having non-physical
momenta in the non-radiative amplitude led some authors
\cite{Lebiedowicz:2021byo, Lebiedowicz:2023mlz, Lebiedowicz:2023ell}
to question the validity of all known formulations of the theorem and to
propose a
modified version.
In this paper, we argue that such criticism has no valid foundation by
explicitly showing that the formulation in \cite{Lebiedowicz:2021byo}
is equivalent at NLP to the ones previously derived in the literature. More
generally,
we
prove the invariance of the LBK theorem at NLP under a specific transformation
of the non-radiative amplitude, which leads to many
equivalent formulations that differ by NNLP corrections.
As a consequence,
the
ambiguities
contained in the LBK theorem due to violation of momentum conservation in the
non-radiative amplitude are power-suppressed beyond the formal validity
of the theorem.

Besides the issue of evaluating the non-radiative processes
using physical on-shell momenta,
other technical aspects must be addressed in a numerical implementation.
In fact, the integration over phase space becomes unstable in the soft limit.
To overcome these instabilities,
the numerical results of this work have been generated with a program
specifically targeted to treat these extreme phase-space configurations.
In addition,
to obtain NLP predictions for an arbitrary process
that can be compared with experimental data,
one wishes to calculate the non-radiative amplitude
using general-purpose event generators.
Thus, in this work, we demonstrate that with our modified shifted momenta
it is possible to obtain predictions for the radiative amplitude
in the soft-photon limit,
using non-radiative amplitudes that are automatically generated.

The structure of this paper is as follows. In \cref{sec:LBK} we
discuss the LBK theorem in all formulations that will be relevant for the
numerical implementation: the one with derivatives, the one with unmodified
shifts and the one with modified shifts. In doing so, we thoroughly
analyse the ambiguities in the computation of the non-radiative process
when the theorem is expressed through derivatives of the non-radiative
amplitude.
 \cref{sec:results}
 contains numerical results for the $e^+e^-\to \mu^+\mu^-\gamma$
and $pp\to \mu^+\mu^-\gamma$ processes. Specifically, after comparing numerical
results based on the
aforementioned
three versions of the LBK theorem, we study the predictive power of the soft
approximation at LP and NLP  in various kinematic ranges. We
conclude in \cref{sec:concl} with a brief discussion.

\section{LBK theorem and shifted kinematics}
\label{sec:LBK}
We begin this section with a compact
review of known results on the LBK theorem. More specifically, in
\cref{sec:trad} we recall the traditional form of the theorem in terms of
derivatives of the non-radiative amplitude, while in \cref{sec:shifts} we
recall the equivalent form of the theorem with shifted kinematics, recently
introduced in \cite{DelDuca:2017twk} and \cite{Bonocore:2021cbv}.
The reason for reviewing these known
forms of the theorem (apart from the sake of comprehensibility and the need to
fix the notation) is twofold. On the one hand, we discuss how an intrinsic
ambiguity in the computation of non-radiative processes does not invalidate the
traditional formulation of the theorem, which has been recently questioned
\cite{Lebiedowicz:2021byo, Lebiedowicz:2023mlz, Lebiedowicz:2023ell}. On
the other hand, we want to stress the virtue of shifting the kinematics, which
removes such ambiguity by restoring momentum conservation. We then
present a new formulation of the theorem in
\cref{sec:modified}, where the shifts are modified in order to keep the
external lines
on the mass shell.

\subsection{Traditional LBK formulation}
\label{sec:trad}

\begin{figure}[tbh]
    \centering
    \begin{subfigure}[b]{0.3\textwidth}
        \centering
            \[
        \vcenter{\hbox{\begin{tikzpicture}
                \begin{feynman}
                \vertex[blob, minimum size=0.75cm] (m) at ( 0, 0) {\(H\)};
                \vertex (a) at (1.2,-0.71);
                \vertex (gi) at (-0.53,0.53);
                \vertex (gf) at (0.71,0.9);
                \vertex (a1) at (0.71,-0.71);
                \vertex (a2) at (0.924,-0.383);
                \vertex (a3) at (1,0);
                \vertex (a4) at (0.924,0.383);
                \vertex (a5) at (0.71,0.71);
                \vertex (b) at (1.2,0.71);
                \vertex (c) at (-1.2,0.71);
                \vertex (b1) at (-0.71,0.71);
                \vertex (b2) at (-0.924,0.383);
                \vertex (b3) at (-1,0);
                \vertex (b4) at (-0.924,-0.383);
                \vertex (b5) at (-0.71,-0.71);
                \vertex (d) at (-1.2,-0.71);
                \vertex (e) at (0, -0.32);
                \diagram* {
                    (a1) -- (m) -- (gi) -- (b1),
                    (a2) -- (m) -- (b2),
                    (a3) -- (m) -- (b3),
                    (a4) -- (m) -- (b4),
                    (a5) -- (m) -- (b5),
                    (gi) --[white, photon,
                    momentum={[arrow distance=1.2mm, arrow style=white] \(k\)}
                    ] (gf),
                };
                \end{feynman}
                \draw[decoration={brace}, decorate]
                (d.south west)-- (c.north west)node[pos=0.5, left] {  \(N\)};
                \draw[decoration={brace}, decorate]
                (b.north east)-- (a.south east)node[pos=0.5, right] {  \(M\)};
                \end{tikzpicture}}}
        \]
        \caption{}
    \end{subfigure}%
    \begin{subfigure}[b]{0.3\textwidth}
        \centering
        \[
        \vcenter{\hbox{\begin{tikzpicture}
                \begin{feynman}
                \vertex[blob, minimum size=0.75cm] (m) at ( 0, 0) {\(H\)};
                \vertex (a) at (1.2,-0.71);
                \vertex (gi) at (-0.53,0.53);
                \vertex (gf) at (0.71,0.9);
                \vertex (a1) at (0.71,-0.71);
                \vertex (a2) at (0.924,-0.383);
                \vertex (a3) at (1,0);
                \vertex (a4) at (0.924,0.383);
                \vertex (a5) at (0.71,0.71);
                \vertex (b) at (1.2,0.71);
                \vertex (c) at (-1.2,0.71);
                \vertex (b1) at (-0.71,0.71);
                \vertex (b2) at (-0.924,0.383);
                \vertex (b3) at (-1,0);
                \vertex (b4) at (-0.924,-0.383);
                \vertex (b5) at (-0.71,-0.71);
                \vertex (d) at (-1.2,-0.71);
                \vertex (e) at (0, -0.32);
                \diagram* {
                    (a1) -- (m) -- (gi) -- (b1),
                    (a2) -- (m) -- (b2),
                    (a3) -- (m) -- (b3),
                    (a4) -- (m) -- (b4),
                    (a5) -- (m) -- (b5),
                    (gi) --[photon, momentum={[arrow distance=1.2mm] \(k\)}]
                    (gf),
                };
                \end{feynman}
                \draw[decoration={brace}, decorate]
                (d.south west)-- (c.north west)node[pos=0.5, left] {  \(N\)};
                \draw[decoration={brace}, decorate]
                (b.north east)-- (a.south east)node[pos=0.5, right] {  \(M\)};
                \end{tikzpicture}}}
        \]
        \caption{}
    \end{subfigure}
    \begin{subfigure}[b]{0.3\textwidth}
    \centering
        \[
    \vcenter{\hbox{\begin{tikzpicture}
            \begin{feynman}
            \vertex[blob, minimum size=0.75cm] (m) at ( 0, 0) {\(H\)};
            \vertex (a) at (1.2,-0.71);
            \vertex (gi) at (-0,0.38);
            \vertex (gf) at (0.71,1.1);
            \vertex (a1) at (0.71,-0.71);
            \vertex (a2) at (0.924,-0.383);
            \vertex (a3) at (1,0);
            \vertex (a4) at (0.924,0.383);
            \vertex (a5) at (0.71,0.71);
            \vertex (b) at (1.2,0.71);
            \vertex (c) at (-1.2,0.71);
            \vertex (b1) at (-0.71,0.71);
            \vertex (b2) at (-0.924,0.383);
            \vertex (b3) at (-1,0);
            \vertex (b4) at (-0.924,-0.383);
            \vertex (b5) at (-0.71,-0.71);
            \vertex (d) at (-1.2,-0.71);
            \vertex (e) at (0, 1.4);
            \diagram* {
                (a1) -- (m) -- (b1),
                (a2) -- (m) -- (b2),
                (a3) -- (m) -- (b3),
                (a4) -- (m) -- (b4),
                (a5) -- (m) -- (b5),
                (gi) --[photon, momentum={[arrow distance=1.2mm] \(k\)}] (gf),
            };
            \end{feynman}
            \draw[decoration={brace}, decorate]
            (d.south west)-- (c.north west)node[pos=0.5, left] {  \(N\)};
            \draw[decoration={brace}, decorate]
            (b.north east)-- (a.south east)node[pos=0.5, right] {  \(M\)};
            \end{tikzpicture}}}
    \]
    \caption{}
\end{subfigure}
    \caption{Diagram (a) corresponds to the non-radiative amplitude $\cal H$,
        where the hard blob $H$ represent an unknown hard interaction. Diagrams
        (b)
        and (c) represent respectively the external and internal contribution
        to
        the radiative amplitude $\cal A$.}
        \label{fig:NMgamma}
\end{figure}
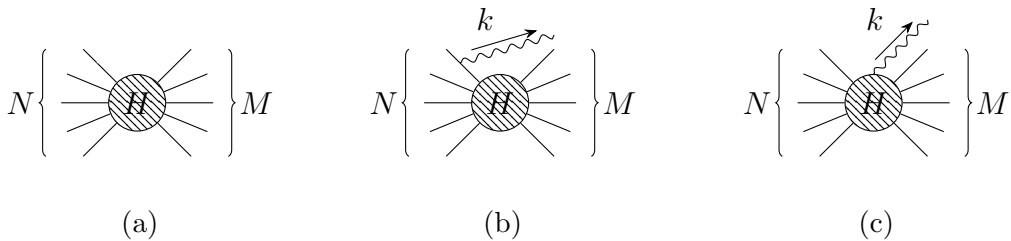

    We consider a generic scattering amplitude ${\cal H}(p_1,\dots,p_n)$
where $N$
particles of hard momenta $p_1,\dots,p_N$ scatter into $M$ particles of hard
momenta
$p_{N+1},\dots,p_{N+M}$, with $n=N+M$.
The
particles interact via an unspecified hard dynamics which can be represented
diagrammatically by the dashed
blob ${H}(p_1,\dots,p_n)$, as in \cref{fig:NMgamma}. For spinning particles $H$
is equal to the full
scattering amplitude ${\cal H}$ stripped off of the external-state wave
functions, while for scalars one trivially has ${\cal H} =H$.

In the radiative process $N\to M+\gamma$, the bremsstrahlung
amplitude
${\cal A}(p_1,\dots,p_n,k)$ includes a photon of momentum $k$ in the final
state.
For reasons that
will become clearer in the next sections, it is convenient to
introduce
a parameter $\eta=1$ for initial particles ($1 \leq i\leq N$) and $\eta=-1$
for
final particles ($N<i\leq n$), so that
momentum conservation reads
\begin{equation}
\sum_{i=1}^{N}p_i - \sum_{i=N+1}^{N+M} p_i = \sum_{i=1}^n \eta_i p_i = k~.
\label{momcons}
\end{equation}
In this way, momenta are incoming for particles in the initial states and
outgoing for particles in the final states.
We also denote with $Q_i$ the
charge of the $i$-th particle.
In the following, we assume that
the momenta $p_i$ appearing both in the non-radiative (i.e. elastic) amplitude
${\cal H}(p_1,\dots,p_n)$ and in the radiative (i.e. inelastic) ${\cal
A}(p_1,\dots,p_n,k)$ fulfill momentum conservation in the radiative
configuration, as in \cref{momcons}. We will
discuss this aspect in detail in \cref{sec:validity}.

The radiative process can be represented
diagrammatically by two classes of
diagrams, as depicted
in
\cref{fig:NMgamma}. In the first one,
the emitted
photon is attached to one of
the external lines. In the second one, the photon couples
directly to some internal line of the unspecified hard subdiagram ${H}$. We
denote the two corresponding radiative amplitudes
(stripped off of
the photon polarization vector $\epsilon_{\mu}(k)$) as
${\cal A}_{\text{ext}}^{\mu}$ and ${\cal A}_{\text{int}}^{\mu}$,
respectively.
We begin with the former.

Without loss of generality, we restrict the analysis to the case of an external
emission
from an initial-state fermion-antifermion pair of charge $Q_1=-Q_2\equiv
Q$ and
momenta $p_1$
and $p_{2}$,
respectively, satisfying $p_1 + p_2 = k$.
The sum of the diagrams corresponding to the two emissions reads
\begin{align}
 {\cal A}_{\text{ext}}^{\mu}(p_1,p_2,k)
= & \bar v( p_2){H}(p_1-k,p_2)
\frac{i(\slashed{p}_1-\slashed{k}+m)}{(p_1-k)^2-m^2}(-iQ
\gamma^\mu)u(p_1)\notag \\
&
 +
\bar v(p_2) (-iQ \gamma^\mu)
\frac{i(
 -\slashed{p}_2 + \slashed{k}
+m)}{(p_2-k)^2-m^2}
{H}(p_1,p_2-k)
u(p_1)~.
\label{toBeExp}
\end{align}
In the limit where the photon momentum $k$ is small compared to the hard
momenta $p_1$ and $p_2$, we can expand\footnote{The expansion in the
four-momentum $k$ is
equivalent to the expansion in the photon energy $\omega$ since all components
of $k$ scale
homogeneously in the soft limit.} in $k$ both the fermion
propagators and the hard subdiagram ${H}$.
 After using the Dirac equation,
enforcing the on-shell
condition $k^2=0$ and neglecting terms proportional to $k^{\mu}$ that vanish by
gauge invariance, we get
\begin{align}
    {\cal A}_{\text{ext}}^{\mu}(p_1, p_2, k) &
    = Q \bar v(p_2){H}(p_1, p_2) \left(
        -\frac{p_1^{\mu}}{p_1 \cdot k}
        + \frac{p_{2}^{\mu}}{p_{2} \cdot k}
        + \frac{ik_{\nu}S^{\mu\nu}}{p_1 \cdot k}
    \right) u(p_1)\notag\\
    &\quad + Q \bar v(p_2) \frac{ik_{\nu}S^{\mu\nu}}{p_{2} \cdot k}
    {H}(p_1, p_2) u(p_1)
    + Q\frac{p^{\mu}_1}{p_1 \cdot k} k^{\nu} \bar v(p_{2})
    \frac{\partial {H}(p_1, p_2)}{\partial p_1^{\nu}} u(p_1) \notag \\& \quad
    - Q\frac{p^{\mu}_{2}}{p_{2} \cdot k} k^{\nu} \bar v(p_{2})
    \frac{\partial {H}(p_1, p_2)}{\partial p_{2}^{\nu}} u(p_1)
    \,\, + \,\,{\cal O}(k)~,
\label{ext}
\end{align}
where we defined $S^{\mu\nu} = \frac{i}{4}\comm{\gamma^\mu}{\gamma^\nu}$
and we exploited the functional dependence of ${H}$ by setting
\begin{align}
\frac{\partial {H}(p_1-k,p_2)}{\partial
    k^{\nu}}\Big|_{k=0}
=-\frac{\partial {H}(p_1,p_2)}{\partial
    p^{\nu}_1}~.
\end{align}

At this point we should note that in order
    to derive \cref{ext}  we
    have expanded \cref{toBeExp} in $k$ while keeping all the other momenta
    fixed, in analogy with Low's derivation in \cite{Low:1958sn}.
    Mathematically, one
    can regard
    the r.h.s. of \cref{toBeExp} as a
    function defined on the entire space spanned by the vectors
    $\{p_1,p_2,k\}$, where
    the vectors
    $p_1$ and
    $p_2$ are not restricted to the surface $p_1+p_2=k$. Although the result of
    such expansion is then defined for arbitrary momenta $p_1$ and $p_2$,
    eventually we
    are of
    course only
    interested in the physical value of the expanded function on the
    momentum-conservation surface.
    An alternative approach, followed e.g. by Burnett and Kroll
    \cite{Burnett:1967km} and more recently by\footnote{We thank O. Nachtmann
    for
        clarifying that in \cite{Lebiedowicz:2021byo,
            Lebiedowicz:2023mlz, Lebiedowicz:2023ell} this is how the expansion
        is performed.}
    \cite{Lebiedowicz:2021byo,
        Lebiedowicz:2023mlz, Lebiedowicz:2023ell}, consists of expanding
    \cref{toBeExp} on the momentum-conservation surface by inserting
    a dependence over $k$ in the momenta
    $p_1(k)$ and $p_2(k)$. The parametrization of the momenta $p_1(k)$ and
    $p_2(k)$, which are not fixed, is then only constrained by
    $p_1(k)+p_2(k)=k$. The two approaches yield
            equivalent expressions on the momentum-conservation constraint up
            to
            power-corrections in the expansion.

To proceed further, one has to compute the
internal
emission contribution ${\cal A}_{\text{int}}$. However,
since in general one cannot know how the photon couples to the internal hard
subdiagrams, one is seemingly prevented from an explicit calculation of ${\cal
A}_{\text{int}}$. However, gauge invariance comes
to the rescue, since
\begin{align}
k_{\mu}\left(
 {\cal A}_{\text{ext}}^{\mu} + {\cal A}_{\text{int}}^{\mu}
\right)=0~.
\end{align}
From this, we deduce that
\begin{align}
    {\cal A}_{\text{int}}^{\mu}&
    = -\sum_{i=1}^2 Q_i \bar{v}(p_2)
    \frac{\partial{H}(p_1, p_2)}{\partial p_{i\,\mu}} u(p_1)
    + K^{\mu}~,
\label{int}
\end{align}
where $K^\mu$ is a gauge invariant term ($k\cdot K=0$). A power
counting analysis reveals that at the tree level $K^{\mu}$ is power-suppressed
at NLP and can be set to zero.\footnote{See \cite{Gervais:2017yxv}
for a more detailed analysis.}

Therefore, combining \cref{ext} and \cref{int} we get the final form for the
LBK theorem for the radiative amplitude ${\cal A}(p_1,p_2,k)$, which reads
\begin{align}
    {\cal A}(p_1,p_2,k)
    =& -\epsilon_{\mu}(k) \sum_{i=1}^2 Q_i
    \,\frac{p_i^{\mu}}{p_i \cdot k}\, {\cal H}(p_1, p_2)
    - \epsilon_{\mu}(k) \sum_{i=1}^2 Q_i \,\bar v(p_2) G^{\mu\nu}_i\,
    \frac{\partial{H}(p_1, p_2)}{\partial p_i^{\nu}} u(p_1) \notag \\&
    + \epsilon_{\mu}(k) Q \,\bar v(p_2)\left[
        {H}(p_1, p_2)\frac{ik_{\nu}S^{\mu\nu}}{p_1\cdot k}
        + \frac{ik_{\nu}S^{\mu\nu}}{p_2\cdot k} {H}(p_1 ,p_2)
    \right] u(p_1)~,
\label{ext+int}
\end{align}
where we have introduced the following tensor
\begin{align}
G^{\mu\nu}_i=g^{\mu\nu}-\frac{(2p_i-k)^{\mu}k^{\nu}}{2p_i\cdot k}
=g^{\mu\nu}-\frac{p_i^{\mu}k^{\nu}}{p_i\cdot k}+{\cal O}(k)~.
\label{Gtensor}
\end{align}
The first term in \cref{ext+int} represents the well-known LP factorization
in terms of the eikonal factor $p\cdot \epsilon/(p\cdot k)$.
The remaining terms correspond to NLP corrections.

A generalization of the calculation above to an arbitrary number of initial or
final state particles is straightforward, although the final result is not
quite compact since one has to distinguish the four cases where the
(anti-)fermion is in the initial or final state.
A short-hand notation that is quite common in the literature on scattering
amplitudes \cite{Cachazo:2014fwa, Strominger:2017zoo,
    Casali:2014xpa, Bern:2014oka,
    Larkoski:2014bxa,
    Luo:2014wea, He:2014bga}
consists of
 factoring out the spin generator and the derivatives from the non-radiative
amplitude, yielding
\begin{align}
{\cal A}(p_1,\dots,p_n,k)
&=\left({\cal S}_{\text{LP}}+{\cal S}_{\text{NLP}}\right){\cal
H}(p_1,\dots,p_n)~,
\label{next-to-soft}
\end{align}
where
\begin{align}
    {\cal S}_{\text{LP}}
    = -\sum_{i=1}^n \eta_i\,Q_i\, \frac{p_i \cdot \epsilon(k)}{p_i \cdot k}~,
    \qquad
    {\cal S}_{\text{NLP}}
    = - \sum_{i=1}^n \eta_i\,Q_i\,
    \frac{ik_{\nu} J_i^{\mu\nu} \epsilon_{\mu}(k)}{p_i \cdot k}~.
    \label{SLP}
\end{align}
Here,
$J^{\mu\nu}_i=S^{\mu\nu}_i+L^{\mu\nu}_i$ is the total angular momentum, while
$L^{\mu\nu}_i=i\left(p_i^{\mu}\frac{\partial}{\partial p_{i\nu}}
-p_i^{\nu}\frac{\partial}{\partial p_{i\mu}}\right)$ is the orbital angular
momentum which is related to the tensor $G^{\mu\nu}$ via
$G^{\mu\nu}_i\,\frac{\partial}{\partial p_i^{\nu}}
=i\frac{k_{\nu}}{p_i\cdot k}L^{\mu\nu}_i$.
However, one should not be fooled by the simplicity of \cref{next-to-soft},
since $J^{\mu\nu}$ is not a simple
multiplicative factor but rather an operator that contains derivatives and
gamma matrices. The derivatives act on the hard
coefficient ${H}$ only (not the full amplitude $\cal H$), while the spin
generator must be inserted in
the
correct
order within the spinors, as shown in \cref{ext+int} for the simple case of an
initial state fermion-antifermion pair.

Things become much simpler for the squared unpolarized amplitude
$\overline{|{\cal A}|}^2$, since all NLP
corrections can be recast in terms of derivatives of the squared non-radiative
amplitude, as first shown in \cite{Burnett:1967km}. This can be seen again by
considering the simple case of two charged
incoming particles as in \cref{ext+int}, where NLP corrections correspond to a
derivative contribution (second term) and a spin contribution (third and fourth
term).
When
squaring and averaging over the polarizations, the non-radiative amplitude
reads simply
\begin{align}
\overline{|{\cal H}(p_1,p_2)|}^2
&=\Tr\left[
 { (\slashed{p}_2 - m) }
{H}(p_1, p_2)
 { (\slashed{p}_1 + m) }
\overline{H}(p_1, p_2))
\right]~,
\end{align}
where we defined $\overline{H}=\gamma^0{H}^{\dag}\gamma^0$.
For the radiative amplitude instead one has the following schematic structure
at NLP
\begin{align}
 {|{\cal A}(p_1,p_2,k)|}^2
&=|{\cal S}_{\text{LP}}|^2 {|{\cal H}(p_1,p_2)|}^2
+2\Re\left(
{\cal S}_{\text{LP}}{\cal H}(p_1,p_2){\cal S}_{\text{NLP}}^{\dagger}
{\cal
H}^{\dagger}(p_1,p_2)
\right)
\label{LP+int}
~,
\end{align}
where ${\cal S}_{\text{LP}}$ and ${\cal S}_{\text{NLP}}$ have been defined in
\cref{next-to-soft}.
The second term in \cref{LP+int} corresponds to
the
interference between the LP factor and either the derivative or the spin
contribution as in \cref{ext+int}.
For an emission from the leg with momentum $p_1$, the spin term becomes
\begin{align}
\Tr\left[
(\slashed{p}_2-m){H}(p_1,p_2)
\left(
\frac{\slashed{k}\gamma_{\mu}}{
 {2}
p_1\cdot k}(\slashed{p}_1+m)
+(\slashed{p}_1+m)\frac{\gamma_{\mu}\slashed{k}}{
 {2}
p_1\cdot k}
\right)
\overline{H}(p_1,p_2))
\right]
\sum_i\eta_i
 {Q_i}
\frac{p_i^{\mu}}{p_i\cdot k}
~.\label{interf}
\end{align}
Up to terms proportional to $k^{\mu}$ one then has
\begin{align}
\frac{\slashed{k}\gamma^{\mu}}{
 {2}
p_1\cdot k}(\slashed{p}_1+m)
+(\slashed{p}_1+m)\frac{\gamma^{\mu}\slashed{k}}{
 {2}
p_1\cdot k}
\,\,=\,\,-\gamma^{\mu}+\frac{p_1^{\mu}}{p_1\cdot k}\slashed{k}
\,\,=\,\,
-G^{\mu\nu}_1\frac{\partial}{\partial p^{\nu}_1}(\slashed{p}_1+m)
~.
\label{ident}
\end{align}
Recalling that derivatives in \cref{ext+int} act on the hard function only,
we conclude that both the spin and the orbital contribution combine into
derivatives of the full squared non-radiative amplitude
$\overline{|{\cal H}(p_1,p_2)|}^2$.
Hence we obtain
\begin{align}
    \savg{{\cal A}(p_1, p_2,k)} &
    = Q^2 \sum_{ij=1}^2 \frac{p_i \cdot p_j}{(p_i \cdot k) (p_j \cdot k)}
    \savg{{\cal H}(p_1, p_2)}
    + Q^2 \sum_{ij=1}^2 \frac{p_{i\,\mu}}{p_i \cdot k} G_j^{\mu\nu}
    \frac{\partial}{\partial p_j^{\nu}} \savg{{\cal H}(p_1, p_2)}~.
\label{squared}
\end{align}

Although so far we have considered fermions, the result above holds also in the
case of spin $0$ and spin $1$ charged particles. In the former case, the spin
generator vanishes, hence it is obvious that NLP terms include only the
derivative contribution. For spin $1$ one can exploit the gauge invariance of
the amplitude to set $\sum_{\lambda}
\epsilon_{\mu}^{(\lambda)}(k)\epsilon_{\nu}^{(\lambda)}(k)=-g_{\mu\nu}$, which
does
not
depend on any momenta and therefore $\frac{\partial}{\partial p}|{\cal
H}|=\frac{\partial}{\partial p}|{H}|$, leaving again only a derivative
contribution.

Finally, we note that \cref{squared} can be trivially generalized to an
arbitrary number $n$ of external (charged or neutral) particles. One  simply
has to repeat the
derivation above for each particle-antiparicle pair, paying special care
to
whether the particles are in the initial or final states.
Thus, the general
form
for the LBK theorem in the traditional formulation reads
\begin{align}
    \savg{{\cal A}(p_1, \dots, p_n,k)}
    =& -\sum_{ij=1}^{n} \eta_i\eta_jQ_iQ_j
    \frac{p_i \cdot p_j}{(p_i \cdot k) (p_j \cdot k)}
    \savg{{\cal H}(p_1, \dots, p_n)} \notag \\&
    - \sum_{ij=1}^{n} \eta_iQ_iQ_j \frac{p_{i\,\mu}}{p_i \cdot k}
    G_j^{\mu\nu} \frac{\partial}{\partial p_j^{\nu}}
    \savg{{\cal H}(p_1,\dots,p_n)}~,
\label{squaredn}
\end{align}
where we used $\eta_j^2=1$.
In \cref{sec:shifts} and \cref{sec:modified} we will discuss two
alternative forms of the theorem that do not involve derivatives.
Before doing so, in the next section we analyse an important property of
\cref{squaredn}.

\subsection{Non-radiative amplitude and unphysical momenta}
\label{sec:validity}

In the traditional form
of the theorem of
\cref{squaredn}, the
non-radiative amplitude\\
${\cal H}(p_1,\dots,p_n)$ is affected by an ambiguity
related to the fact that it must be evaluated outside the physical region.
Indeed, in order for ${\cal H}(p_1,\dots,p_n)$
to represent a physical process with no photon radiation,
the
momenta $p_i$ should fulfill $\sum_i \eta_ip_i=0$. However, the momenta $p_i$
that have been introduced in \cref{momcons} fulfill momentum conservation in the
radiative amplitude ${\cal A}(p_1,\dots,p_n, k)$, i.e. $\sum_i\eta_ip_i=k$.
Therefore,
in \cref{squaredn} we are in fact evaluating
$\mathcal{H}$
using radiative momenta, which for $k\neq 0$ are unphysical for the
non-radiative
process and thus induce an unphysical ambiguity in the final result.
It is the aim of this section to demonstrate that
the use of unphysical  momenta in the non-radiative amplitude does not
invalidate the
 consistency of
\cref{squaredn} at LP and NLP.

We start with the observation that every amplitude, and in particular the
non-radiative amplitude of the LBK theorem $\mathcal{H}(p_1, \ldots,
p_n)$, is intrinsically ambiguous if momentum conservation is not imposed.
In fact, one can always find a
function $\Delta$ such that the transformation
\begin{equation}
{\mathcal{H}}(p_1, \ldots, p_n) \to \mathcal{H}(p_1, \ldots, p_n) +
\Delta(p_1,\ldots, p_n)
\label{delta}
\end{equation}
leads to the exact same physics, as long as $\Delta$ fulfills
\begin{equation}
\Delta(p_1,\ldots, p_n)\,\delta\left(\sum_i \eta_i p_i\right)=0~.
\label{delta2}
    \end{equation}
We would like to exploit this property to show that \cref{squaredn} does not
depend on $\Delta$, up to NNLP corrections. To do so, we
have to assign a scaling in $k$ to $\Delta$ in \cref{delta}.
In this regard, we note that momentum conservation for the radiative
amplitude
must be fulfilled in order for \cref{squaredn} to give physical
results.
Therefore, what matters for the invariance of \cref{squaredn} under
\cref{delta} is the
value of
$\Delta$ on the momentum-conservation surface $\sum_i \eta_i p_i=k$.
By imposing this constraint, the momenta $p_i$ can be effectively interpreted
as
functions
$p_i(k)$, with an arbitrary functional dependence over $k$, constrained only by
total momentum conservation.
This induces an implicit dependence of $\Delta(p_1, \ldots, p_n)$
over $k$ through the momenta $p_i(k)$, such that $\Delta$ can be
expanded in $k$.
Therefore, we can now check whether the r.h.s. of \cref{squaredn} is
invariant at NLP  under the
transformation of
\cref{delta} on the
momentum-conserving surface $\sum_i \eta_i p_i=k$.
Let us consider the LP and NLP cases separately. To simplify the discussion, we
will first consider the form of the LBK theorem at the amplitude level as in
\cref{next-to-soft} and \cref{SLP} in the scalar
case only. We will then discuss how the generalization for the squared
amplitude (which is valid also in the spinning case) follows analogously.

To check whether the LBK
theorem in the form of \cref{next-to-soft} and \cref{SLP} is invariant
under
\cref{delta} at LP, one has to verify that
\begin{equation}
\left(\sum_i \eta_i Q_i
\frac{p^\mu_i(k)}{p_i(k)\cdot k}\right)\Delta(p_1(k), \ldots, p_n(k)) =
\order{1},
\label{LPDelta}
\end{equation}
or alternatively
\begin{align}
\Delta(p_1(k), \ldots, p_n(k))=O(k)~.
\label{deltaLP}
\end{align}
The key point here is to notice that the limit $k\to 0$ implies the
expression $\sum_i \eta_i p_i = 0$. Then, from \cref{delta2}
one concludes that $\Delta\to 0$ for $k\to 0$.
Since in this paper we are restricting the scope of our analysis to a
tree-level calculation, where the absence of non-analytic terms allows a
Laurent expansion in $k$, we
conclude that $\Delta$ is at worst of $\mathcal O(k)$ and hence \cref{deltaLP}
is fulfilled, thus validating the theorem at
LP.

At NLP we have to modify the consistency condition of \cref{LPDelta} as follows
\begin{align}
\sum_i\eta_i Q_i\left[\frac{p_i^\mu(k)}{k\cdot p_i(k)} \Delta(p_1(k), \ldots,
p_n(k))
+\eta_i G_i^{\mu \nu}\pdv{\Delta(p_1(k), \ldots, p_n(k))}{p_{i}^\nu}\right] =
\order{k}
\label{NLPDelta}~,
\end{align}
where $\mathcal O(k)$ in the r.h.s. represent NNLP corrections.
In order to verify this condition, once again we introduce a $k$ dependence
in $\Delta(p_1, \ldots, p_n)$
via $p_i(k)$. Given that we have to deal with derivatives,
it
is convenient to make the dependence on $k$ explicit by
defining a new function $\tilde{\Delta}_{\mu}(p_1, \ldots, p_n, k)$ which
is constrained on the momentum-conservation surface by
\begin{align}
k^{\mu}\tilde{\Delta}_{\mu}(p_1, \ldots, p_n,
k)\,\,\delta\left(\sum_i\eta_ip_i-k\right)
={\Delta}(p_1(k), \ldots, p_n(k)) \,\,\delta\left(\sum_i\eta_ip_i-k\right)~.
\label{surface}
\end{align}
By enforcing the delta constraints of \cref{surface}, one can effectively
substitute
$k=k(p)=\sum_i \eta_i
p_i$ in $k^{\mu}\tilde{\Delta}_{\mu}$. Therefore, the
following
relation between the derivatives of $\Delta$ and $\tilde{\Delta}_{\mu}$ can be
found
\begin{align}
\pdv{\Delta(p_1, \ldots, p_n)}{p_j^\mu} &
=\frac{d}{dp_j^{\mu}}\left(
k^{\nu}(p)\tilde{\Delta}_{\nu}(p_1, \ldots, p_n,
k(p))\right)\notag \\&
=  \eta_j
\tilde{\Delta}_\mu(p_1, \ldots, p_n, k(p))+k^\nu(p)
\pdv{\tilde{\Delta}_\nu(p_1,
\ldots, p_n, k(p))}{p_j^\mu} \notag \\
&\quad  + \eta_j k^\nu(p)
\pdv{\tilde{\Delta}_\nu(p_1, \ldots, p_n, k(p))}{k^\mu}\notag \\&
= \eta_j \tilde{\Delta}_\mu(p_1, \ldots, p_n, k(p)) + \order{k}~,
\end{align}
where in the last equality we dropped terms of order $\order{k}$, using the
fact that $\tilde\Delta_{\mu}=\order{1}$.
At this point \cref{NLPDelta} follows straightforwardly. Indeed, the l.h.s.
of \cref{NLPDelta} becomes
\begin{align}
\sum_i\eta_i Q_i\left[\frac{p_i^\mu}{k\cdot p_i} k^\nu \tilde{\Delta}_\nu(p_1,
\ldots, p_n, k)
+\eta_i \left(g^{\mu\nu}-\frac{p_i^\mu k^\nu}{p_i\cdot k}\right)\eta_i
\tilde{\Delta}_\nu(p_1, \ldots, p_n, k)\right] + \order{k}~.
\label{lhs}
\end{align}
For a given $i$, thanks to \cref{LPDelta} and \cref{surface}, all
terms in \cref{lhs} are NLP. However,
 because
 $\eta_i^2=1$, there is a cancellation between the terms in
\cref{lhs}, yielding
\begin{align}
\left(\sum_i\eta_i Q_i\right)\tilde{\Delta}^\mu(p_1, \ldots, p_n, k) +
\order{k}~.
\label{condit}
\end{align}
Finally, given that $\sum_i\eta_i Q_i=0$ by charge conservation,  only a
residual $\mathcal O(k)$ (i.e. NNLP) term remains, as required by
\cref{NLPDelta}. We then conclude that the NLP theorem at the amplitude level
as in \cref{next-to-soft} and \cref{SLP} is invariant under \cref{delta} on the
surface
$\sum_i\eta_ip_i=k$ and thus it is consistent also when the corresponding
non-radiative
amplitude is evaluated with unphysical momenta.

A crucial step in the derivation above is the cancellation of NLP ambiguities
between the LP term and the derivative term. To make the general
arguments discussed above more concrete and see this cancellation explicitly,
in \cref{sec:comp}
we consider the soft
bremsstrahlung in a simple case of a $2\to 2$ non-radiative process involving
only scalar particles.
This discussion is also meant to clarify the relation with the work of
\cite{Lebiedowicz:2021byo, Lebiedowicz:2023ell, Lebiedowicz:2023mlz}
where the validity of the traditional form of the LBK theorem has been
questioned.

The generalization of the previous arguments to the squared-matrix elements of\\
\cref{squaredn} straightforwardly carries over, by simply adjusting the correct
power counting in $k$. More specifically, one has to check that
\cref{squaredn} remains invariant under\footnote{Note that although the
definition of $\Delta$ here is
different from the one in \cref{delta}, it obeys \cref{delta2}.}\\
$\overline{|{\cal H}|}^2
\to \overline{|{\cal H}|}^2 +\Delta $.
At LP this is equivalent to showing that
\begin{equation}
\left(\sum_{ij} \eta_i\eta_j Q_i Q_j
\frac{p_i(k)\cdot p_j(k)}{p_i(k)\cdot k\,\, p_j(k)\cdot k}\right)\Delta(p_1(k),
\ldots, p_n(k))
= \order{k^{-1}},
\label{LPDeltaSquared}
\end{equation}
while at NLP the consistency condition reads
\begin{align}
&\sum_{ij}\eta_i\eta_j Q_iQ_j\Bigg(\frac{p_i(k)\cdot p_j(k)}{p_i(k)\cdot k\,\,
p_j(k)\cdot k}
\Delta(p_1(k), \ldots, p_n(k))\notag\\
&\quad +\frac{p_{j\,\mu}(k)}{p_j(k)\cdot k}\eta_i G_i^{\mu
\nu}\pdv{\Delta(p_1(k),
\ldots,
p_n(k))}{p_{i}^\nu}\Bigg) =
\order{1}
\label{NLPDeltaSquared}~.
\end{align}
Both conditions can be verified with the same arguments as outlined above,
thus showing that the r.h.s. of \cref{squaredn} does not depend on $\Delta$ at
LP and NLP.
Therefore,
even though the non-radiative
amplitude is evaluated with unphysical momenta, the formulation of the theorem
as in
\cref{squaredn} is consistent at NLP.

Finally, we note that the arbitrariness in the evaluation of the non-radiative
function with non-physical momenta was already observed
by Burnett and Kroll in their original work \cite{Burnett:1967km}.
In fact, Burnett and Kroll proposed a prescription to evaluate the
non-radiative ampitude by shifting the unphysical momenta by an arbitrary
quantity that restore momentum conservation in the elastic amplitude. The
argument we have presented here,
instead, is more general. By exploiting the invariance at NLP of the
non-radiative
amplitude under \cref{delta} we have proven that \cref{squaredn} is
consistent without the need to restore momentum conservation. In fact, one
could restrict the transformations of
\cref{delta} to the special case of linear shifts on the external momenta.
Then, the proposal of Burnett and Kroll would correspond to the specific case
of shifts that fulfill momentum conservation in the elastic configuration.
In order to shed light on the relation between the general argument of this
section and the strategy of Burnett and Kroll, in \cref{sec:amb} we discuss
the invariance of \cref{squaredn} in the special case where
\cref{delta} can be
represented by linear transformations of the momenta.

\subsection{From derivatives to shifts}
\label{sec:shifts}

In the previous section we have verified that the traditional form of the LBK
theorem with derivatives of the non-radiative process is consistent at NLP,
since
non-physical
ambiguities arising in the computation of the non-radiative process are NNLP.
Still, the dependence of \cref{squaredn} on an unphysical
non-radiative amplitude
seems  unsatisfactory.
In particular,
if one intends to automatically generate the amplitude of the
non-radiative process using publicly available tools, having a form of the
theorem that
is defined from scratch for physical amplitudes with momenta that fulfill
momentum
conservation is desirable. The non-radiative process is then
 computed for physical momenta and is thus unambiguous.
 Hence,
it is natural to ask whether it is possible to find a simpler formulation
of the theorem which is particularly suitable for numerical implementations.

The answer is yes, as proposed in \cite{Bonocore:2021cbv},
building on previous work in QCD \cite{DelDuca:2017twk}. It stems from the fact
that since derivatives are the generators of translations, one can convert
the term with derivatives in the LBK theorem into momentum shifts   in the
non-radiative
amplitude. In fact, one can write \cref{squaredn}
as
\begin{align}
    \savg{\mathcal{A}(p_1, \dots, p_n, k)} &
    = \savg{{\cal S}_{\text{LP}}} \left[
        1
        + \sum_j \delta p_j^{\nu} \pdv{p_j^\nu}
    \right] \savg{\mathcal{H}(p_1, \dots, p_n)}~,
\label{matching}
\end{align}
where the shifts $\delta p_i$ are to be determined, while from \cref{SLP}
\begin{equation}
    \savg{{\cal S}_{\text{LP}}}
    = -\sum_{ij=1}^{n} \eta_i \eta_j Q_i Q_j
    \frac{p_i \cdot p_j}{(p_i \cdot k) (p_j \cdot k)}~.
\label{SLPsavg}
\end{equation}
By comparison with \cref{squaredn}, we deduce
\begin{align}
    \delta p_j^\nu &
    = Q_j \left(
        \sum_{k,l} \eta_k \eta_l Q_k Q_l
        \frac{p_k \cdot p_l}{(p_k \cdot k)(p_l \cdot k)}
    \right)^{-1}
    \sum_{i} \left(\frac{\eta_iQ_i p_{i\mu}}{k \cdot p_i}\right) G_j^{\mu\nu}~.
\label{deltapi}
\end{align}
Thus,
one obtains
\begin{align}
    \savg{\mathcal{A}(p_1, \dots, p_n, k)} &
    = \savg{{\cal S}_{\text{LP}}}
    \savg{\mathcal{H}(p_1 + \delta p_1, \dots, p_n + \delta p_n)}~,
\label{LBKshifts}
\end{align}
i.e. a form of the LBK theorem without derivatives and with a single LP soft
factor.

We note immediately that $\delta p_j^\nu= \order{k}$, hence the shifts vanish
at LP, as expected. Another crucial property that can be readily verified
is that
\begin{align}
    \sum_j \eta_j \delta p_j^\mu = -k^\mu~.
\label{deltak}
\end{align}
Therefore, recalling that $\sum_j \eta_j p_j^\mu = k^\mu$, we deduce that
momentum conservation is restored in the non-radiative amplitude of
\cref{LBKshifts},
which can be then computed without the ambiguities discussed
in the previous section.

Note also that by getting rid of the derivatives, we
obtained a form of the
theorem with just a single positive-defined term. Naturally, as long as the
soft expansion is meaningful, we expect the derivative term in
\cref{squaredn} to be small w.r.t. the LP term. Thus, for soft-photon momenta,
also \cref{squaredn} remains
positive, as expected for a cross-section. Still, for a theorem whose scope is
to extend the range of validity of the soft approximation to larger soft
momenta, the formulation in
\cref{LBKshifts} seems more elegant, since it ensures that the cross-section
remains positive. We will
come back to this point in \cref{sec:results}.

Finally, one can easily verify that the momenta shifts are orthogonal to each
momentum, i.e.
\begin{align}
\delta p_j \cdot p_j = 0~.
\end{align}
This implies that
\begin{align}
(p_j+\delta p_j)^2 = m_j^2 + \order{k^2}~,
\label{onshell}
\end{align}
thus fulfilling the on-shell condition up NLP. We notice however that the
condition is
violated already at NNLP. More precisely, one can
verify that
\begin{align}
(\delta p_j)^2 &
=Q^2_j\left(\sum_{k, l} \eta_k \eta_l Q_k Q_l \frac{p_k \cdot p_l}{(p_k \cdot
    k)(p_l \cdot k)}\right)^{-1} \neq 0~,
\end{align}
hence masses do get shifted by a non-zero NNLP
amount for non-vanishing $k$. This feature might be problematic when using
automatically generated amplitudes, since most of the public tools typically
require
momenta to be exactly on-shell. In the next section we discuss how to overcome
this problem.

\subsection{Modified shifted kinematics}
\label{sec:modified}

We seek another expression for $\delta p_i$ that ensures that masses are
not shifted, without spoiling the NLP terms of the LBK theorem. Hence we require
the new definition for $\delta p_i$ to
fulfill the following conditions:
\begin{enumerate}[label=(\roman*)]
    \item it conserves momentum to all orders in $k$, i.e.
    \begin{align}
     \sum_i \eta_i \delta p_i +k=0~,
     \label{eins}
    \end{align}
    \item it fulfills the on-shell condition to all orders in $k$, i.e.
    \begin{align}
    (p_i+\delta p_i)^2 =
    m_i^2~,
    \label{zwei}
    \end{align}
    \item it reduces to \cref{deltapi} up to NNLP corrections, i.e.
    \begin{align}
        \delta p_j^\nu
        = -Q_j \left(\savg{{\cal S}_{\text{LP}}}\right)^{-1}
        \sum_{i} \left(\frac{\eta_i Q_i}{k \cdot p_i}\right)
        \left(p_i^\nu - \frac{p_i \cdot p_j }{p_j \cdot k} k^\nu\right)
        + \order{k^2}~.
    \label{drei}
    \end{align}
\end{enumerate}
We can find such definition by considering the following ansatz,
\begin{align}
    \delta p_i^\mu = \sum_j A_{ij} p_{j}^\mu + B_{i} k^\mu~,
\end{align}
and, by imposing the
constraints (i)-(iii),  subsequently determine the unknown coefficients
$A_{ij}$ and
$B_i$ .
It turns out that these conditions are not too restrictive and one is free to
select a single solution.
Details of this calculation can be found in \cref{sec:ansatz}. The final result
reads
\begin{align}
    \delta p_i^\mu
    = A Q_i \sum_j \frac{\eta_j Q_j}{k \cdot p_j} p_{j\nu} G^{\nu\mu}_i
    + \frac{1}{2} \frac{A^2 Q^2_i {\savg{{\cal S}_{\text{LP}}}}}{p_i \cdot k}
    k^\mu~,
\label{modshifts}
\end{align}
with
\begin{align}
    A
    = \frac{1}{\chi} \left(
        \sqrt{1 - \frac{2\chi}{\savg{{\cal S}_{\text{LP}}}}}
        - 1
    \right)~,
    \qquad
    \chi = \sum_i \frac{\eta_i Q^2_i}{p_i \cdot k}~,
\end{align}
and $\savg{{\cal S}_{\text{LP}}}$ defined in \cref{SLPsavg}.
It is straightforward to check that the conditions of \cref{eins}, \cref{zwei}
and \cref{drei} are satisfied by this solution. Indeed, \cref{modshifts}
reduces to \cref{deltapi} up to NNLP corrections and therefore it still
correctly reproduces the LBK theorem at NLP.
Moreover, both momentum conservation and the on-shell
condition hold to all-orders in the soft
expansion.

The price to pay is that we need to introduce spurious NNLP terms in the hard
momenta, which
unavoidably affect the numerical evaluation of the non-radiative amplitude
$\cal H$,
hence the
prediction for the photon spectra.
However, one should bear in mind that the sensitivity of $\cal H$ to NNLP
effects is not a
feature that belongs only to the modified kinematics. We encountered it also in
the other two versions of the LBK theorem. Specifically, in the traditional
form with
derivatives, $\cal H$ is not uniquely defined.
Thus, even though the NLP ambiguities cancel,
as we showed in \cref{sec:validity}, NNLP spurious terms do survive.
In the formulation with unmodified shifted kinematics,
although there are no unphysical ambiguities
due to violation of momentum conservation,
when going from \cref{matching} to \cref{LBKshifts}
we are implicitly adding spurious NNLP terms. Hence \cref{LBKshifts} is also
valid only up to NLP.

More generally, we note that a residual arbitrariness in the final result due to
missing higher-order terms is a feature common to all perturbative expansions.
In fact, choosing a specific definition for the modified shifts in
\cref{LBKshifts}
corresponds to the choice of a ``scheme'', which is specified by the inclusion
of
power-suppressed (i.e. beyond NLP) terms. In this regard, we note that the
modified shifts make this scheme-dependence transparent, since the choice
is process-independent. Instead, in the traditional formulation of the LBK
theorem of
\cref{squaredn}, $\cal H$ is not univocally determined and thus
the
(hidden) scheme-dependence corresponds to the choice of a specific
functional form for the
amplitude. This choice is obviously process-dependent.

The question that remains is what is the role of these NNLP effects in a
numerical computation of photon spectra, i.e. what is the version of the LBK
theorem that gives the best approximation of the exact radiative process. Among
other
things, we investigate these aspects in the next section.

\section{Numerical predictions for $e^+e^-\to \mu^+\mu^-\gamma$ and $pp\to
\mu^+\mu^-\gamma$}
\label{sec:results}

In the following, we present a numerical study of the spectra of soft
photons produced in association with a muon pair in $e^+e^-$ and $pp$
collisions.  The cross-sections for the  $ij \to \mu^+\mu^- (+ \gamma)$
processes ($ij = e^+e^-, \, q \bar q$) are calculated at the tree level,
including
both $Z$ and $\gamma$ exchange. We consider the $e^+e^-$ collisions at the
center-of-mass (c.m.) energy of 91 GeV, i.e. the LEP1 collision energy, at
which the measurements of photon spectra were carried out by the DELPHI
collaboration\cite{DELPHI:2007nmh}. The $pp$ collisions are considered at 14
TeV c.m. energy. To ensure that we are not
sensitive to any infrared effects other than that related to the soft photon,
we impose kinematical cuts on the transverse momentum of the muons,
$p_{T,\mu}>10$ GeV, pseudorapidity of all final state particles,
$|\eta_i|<2.5$,
as well as the photon-muon separation, $\Delta R>0.4$. The photon distributions
are computed with an in-house code using the VEGAS+
algorithm \cite{Lepage:2020tgj}
for performing the phase-space integration. In the case of $pp$ collisions, we
make use of LHAPDF6\cite{Buckley:2014ana} and choose to evaluate the cross
sections with NNPDF4.0\cite{NNPDF:2021njg} LO set of parton distribution
functions. The tree-level amplitudes for the non-radiative and (exact) radiative
amplitudes are either generated by MadGraph5@NLO\cite{Alwall:2014hca} or
calculated analytically.\footnote{
    The analytical expression for the non-radiative amplitude used in this
    section is specified by
    $\mathcal{H}(p_1, p_2, p_3, p_4) = \mathcal{H}(s(p_1, p_2), t(p_1, p_3)))$.
}
All exact (i.e. obtained without imposing the soft-photon
approximation) results have been cross-checked against numerical results
generated using the SHERPA event generator.\footnote{Numerical checks with
MadGraph5 were also performed. The results of MadGraph5 for the\\
$e^+e^- \to \mu^+ \mu^- \gamma$
process appear to depend on the chosen integration strategy
and the way of grouping the Feynman diagrams for calculations. We thank the
MadGraph team for clarifying that point.}

The exact predictions and the predictions to which we refer to as NLP are
obtained by integrating the exact matrix elements or their particular NLP
approximation over the full 3-particle phase space. At this point, we note that
one could also consider the expansion of the phase-space factor in powers of
the
soft momentum and truncate it at LP or NLP depending on whether the matrix
elements are evaluated at NLP or LP, respectively. However, as discussed in the
last section, the soft approximation of the matrix elements based on the LBK
theorem in all
its forms receives NNLP contributions. Therefore, integrating over
the full phase space leads to the same level of accuracy. On the other hand,
our LP predictions are
obtained by imposing momentum conservation on all external particles other than
the photon, which effectively corresponds to truncating the expansion of the
phase-space factor at LP and calculating the LP term of the non-radiative
amplitude on such external momenta.

\begin{figure}[htb]
    \centering
    \begin{subfigure}[b]{0.495\textwidth}
        \includegraphics[width=\textwidth]{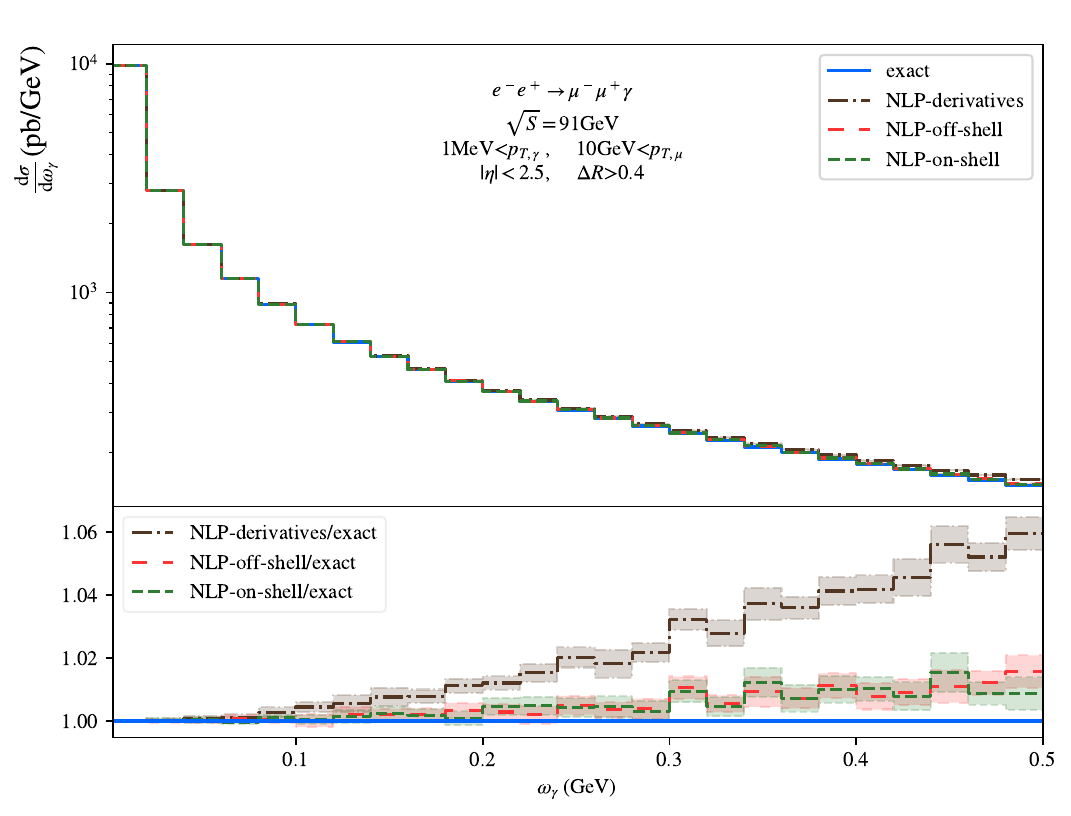}
    \end{subfigure}
    \begin{subfigure}[b]{0.495\textwidth}
        \includegraphics[width=\textwidth]{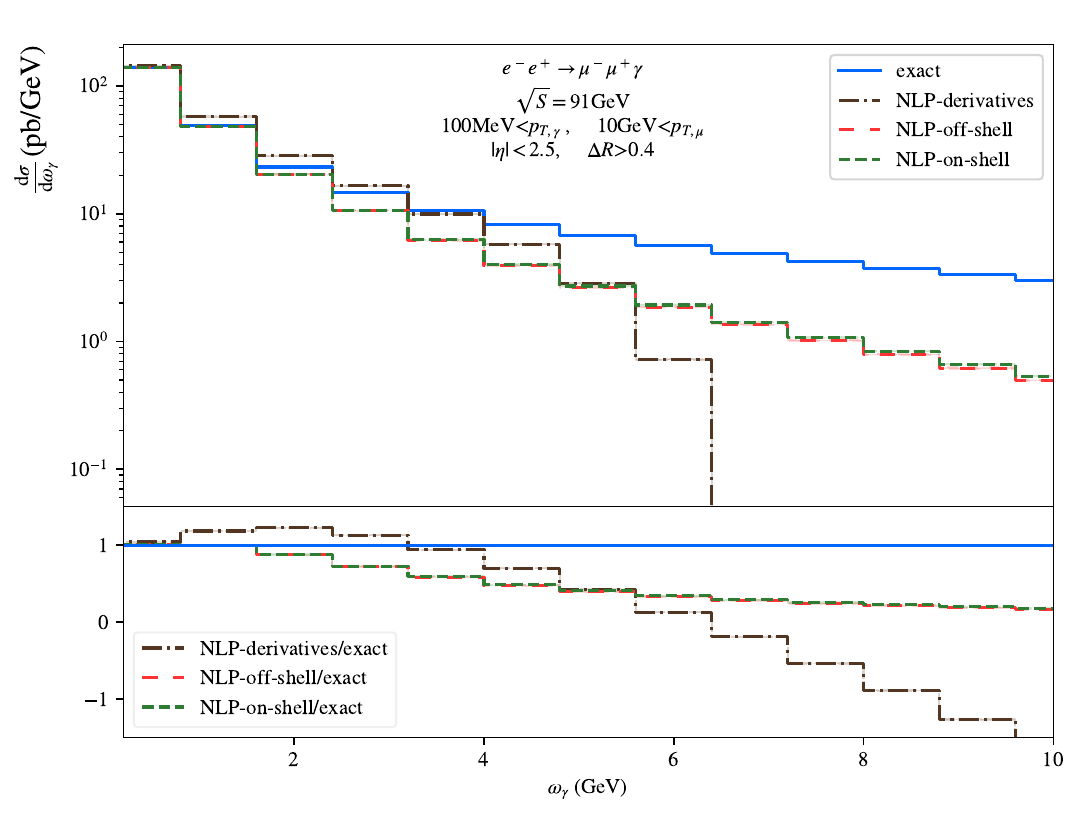}
    \end{subfigure}
    \caption{Comparison of the soft-photon energy spectra calculated using
    three formulations of the LBK theorem discussed in this paper
        with the exact (i.e. no soft expansion) result for the process $e^-e^+
        \to \mu^-\mu^+ \gamma$ at $\sqrt s = 91$ GeV.}
    \label{figNLP}
\end{figure}

We begin with a comparison of numerical predictions for the $e^+e^- \to
\mu^+ \mu^- \gamma$ process obtained using the NLP approximations of the
amplitude derived in the previous chapter, i.e. the traditional form with
    derivatives of \cref{squaredn}, the form of \cref{LBKshifts} with the
    off-shell
    momenta shifts defined in \cref{deltapi}, and the form of \cref{LBKshifts}
    with
    on-shell shifts defined in \cref{modshifts}. To this aim, we use the
    analytic result for the non-radiative amplitude to calculate the
    derivatives in \cref{squaredn} analytically, as well as to compute
    \cref{LBKshifts} involving the off-shell momenta shifts. We also compare
    the NLP predictions to the full result where no soft approximation has been
    applied. The corresponding differential distributions in photon energy
    $\omega$ are shown in \cref{figNLP}, in the range of 1-500 MeV (left plot)
    and 0.1-10 GeV (right plot). As expected, we observe  that all three
    approaches converge to the exact result in the limit of small $\omega$, and
    depart from it with growing photon energy. However, the approximation of
    the exact result provided by the formulation of the LBK theorem involving
    derivatives, \cref{squaredn}, is distinctively worse than those based on
    formulations involving shifting of momenta. While the latter agree with the
    exact result within 1-2\% for $\omega \lesssim 500 \text{MeV}$, where one
    naively expects the soft approximation to work, the former differs from the
    exact predictions within the same range by up to 6\%. Besides, its
    behaviour is qualitatively different: at $\omega \gtrsim 6\text{GeV}$, the
    derivative approach gives a non-physical negative result, in contrast to
    the other predictions which stay positive. This can be understood from
    \cref{squaredn} which is a sum of two not-positive-defined terms and as
    such can get negative when the expansion breaks down. As the difference
    between various NLP approximations is due to the NNLP terms, these results
    clearly show the relevance of the subleading terms beyond the formal
    accuracy of the LBK theorem.

We also see that the NLP approximations of the photon spectrum calculated using
the non-radiative amplitude  with momenta shifted on-shell or off-shell perform
equally well for the photon energies considered here, indicating that the NNLP
effects introduced in the on-shell shifts are not significant. Since the
formulation with momenta shifted on-shell enables sourcing the amplitude
subroutines from a wide range of public tools, we employ this formulation in
further studies.

\begin{figure}[htb]
    \centering
    \begin{subfigure}[b]{0.495\textwidth}
        \includegraphics[width=\textwidth]{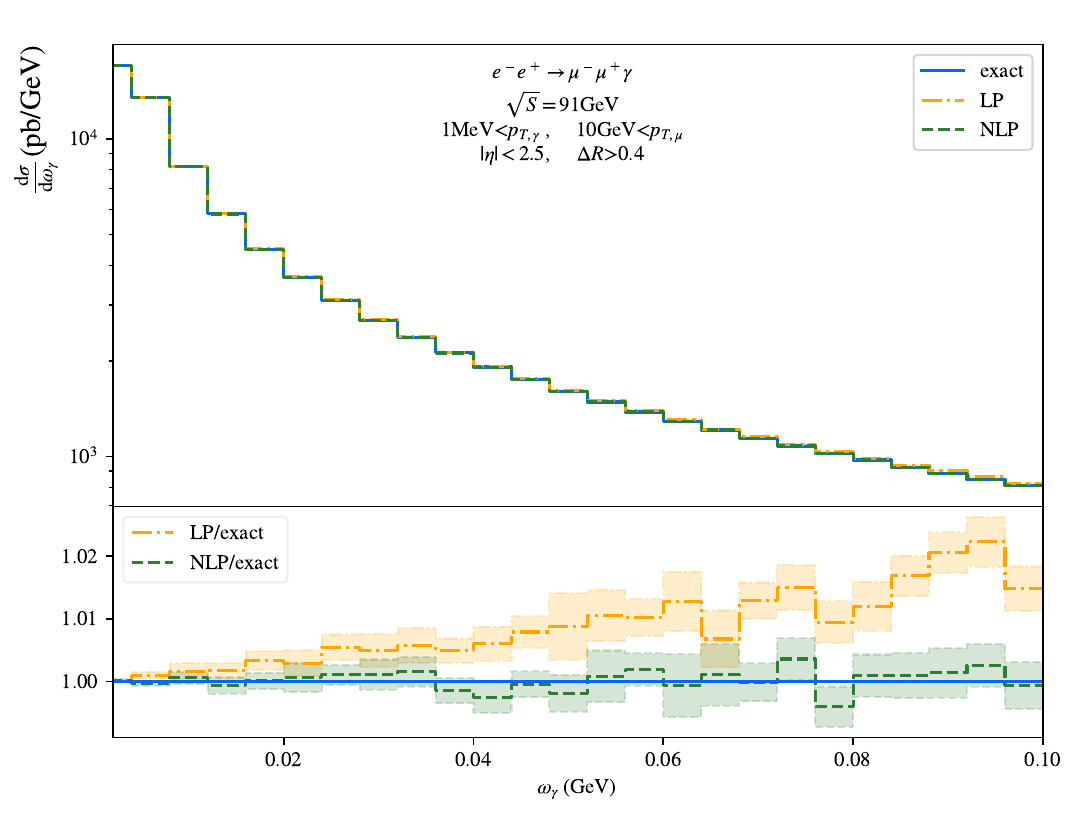}
    \end{subfigure}
    \begin{subfigure}[b]{0.495\textwidth}
        \includegraphics[width=\textwidth]{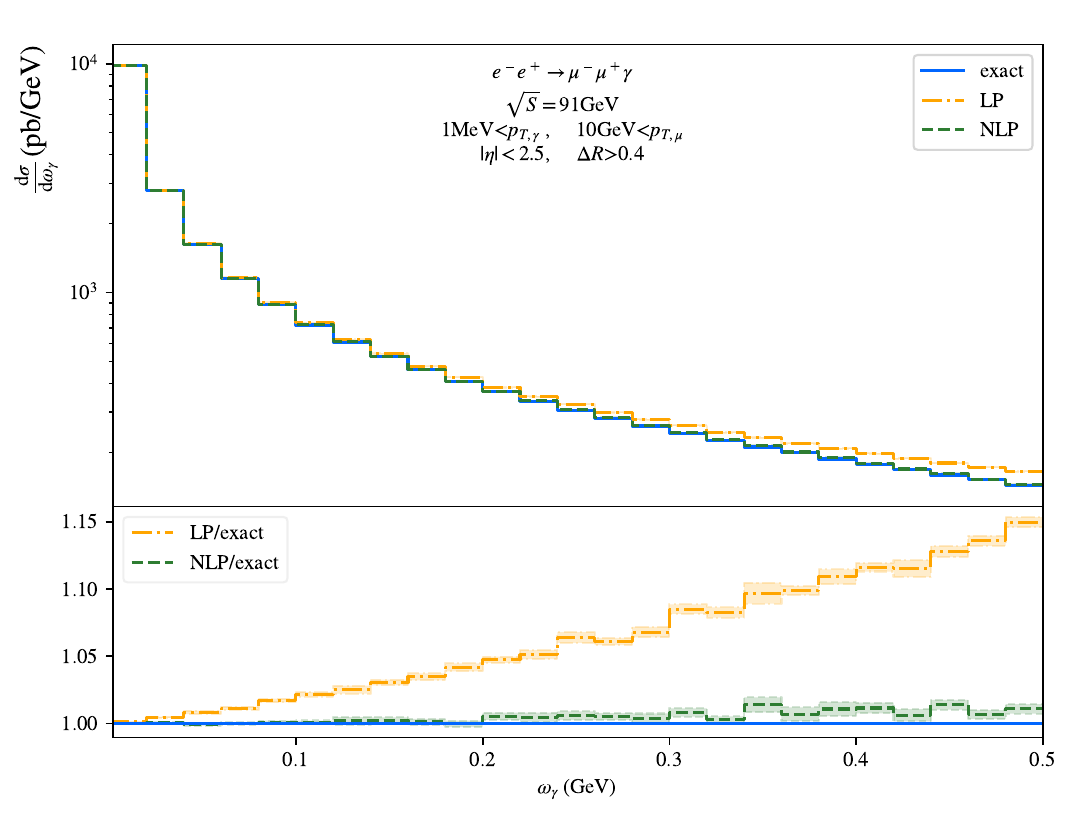}
    \end{subfigure}
    \\
    \begin{subfigure}[b]{0.495\textwidth}
        \includegraphics[width=\textwidth]{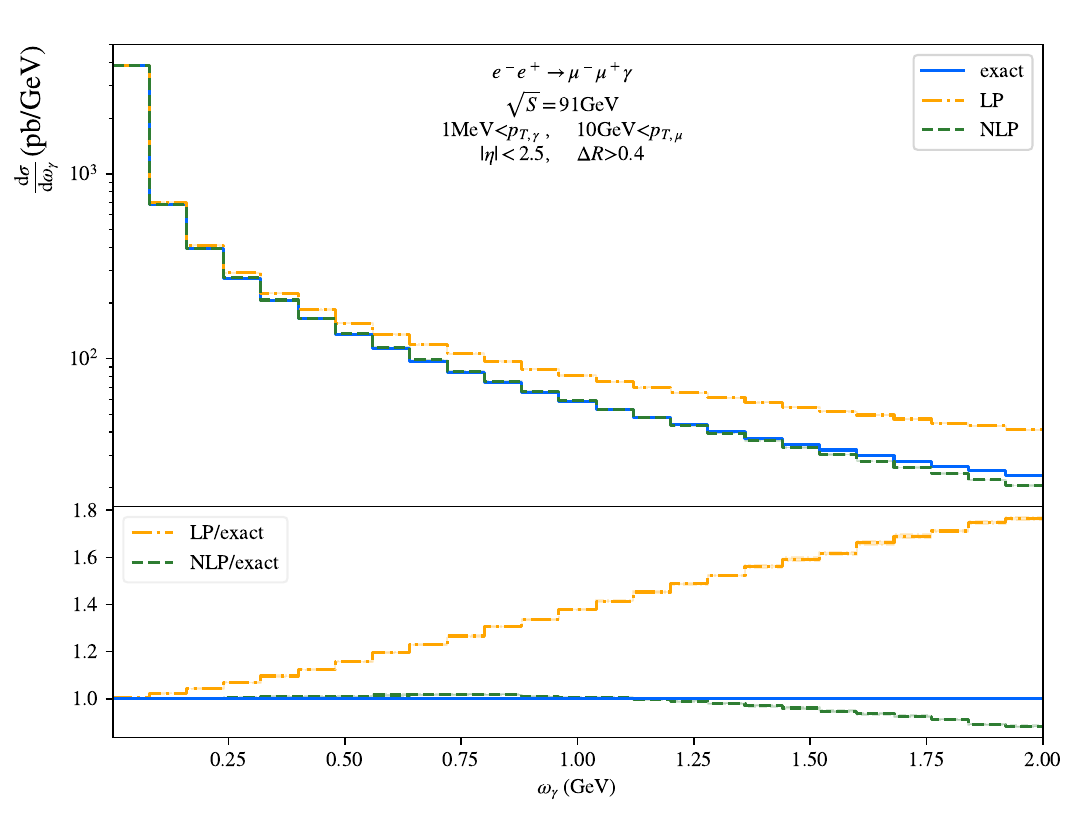}
    \end{subfigure}
    \caption{The $\LP$ and $\NLP$ approximations and the exact result for the
    energy spectrum of the photon in the $e^- e^+
        \to \mu^-
        \mu^+ \gamma$ process at $\sqrt s =91$ GeV.}
    \label{fig:epemomega}
\end{figure}

\begin{figure}[htb]
    \centering
    \begin{subfigure}[b]{0.495\textwidth}
        \includegraphics[width=\textwidth]{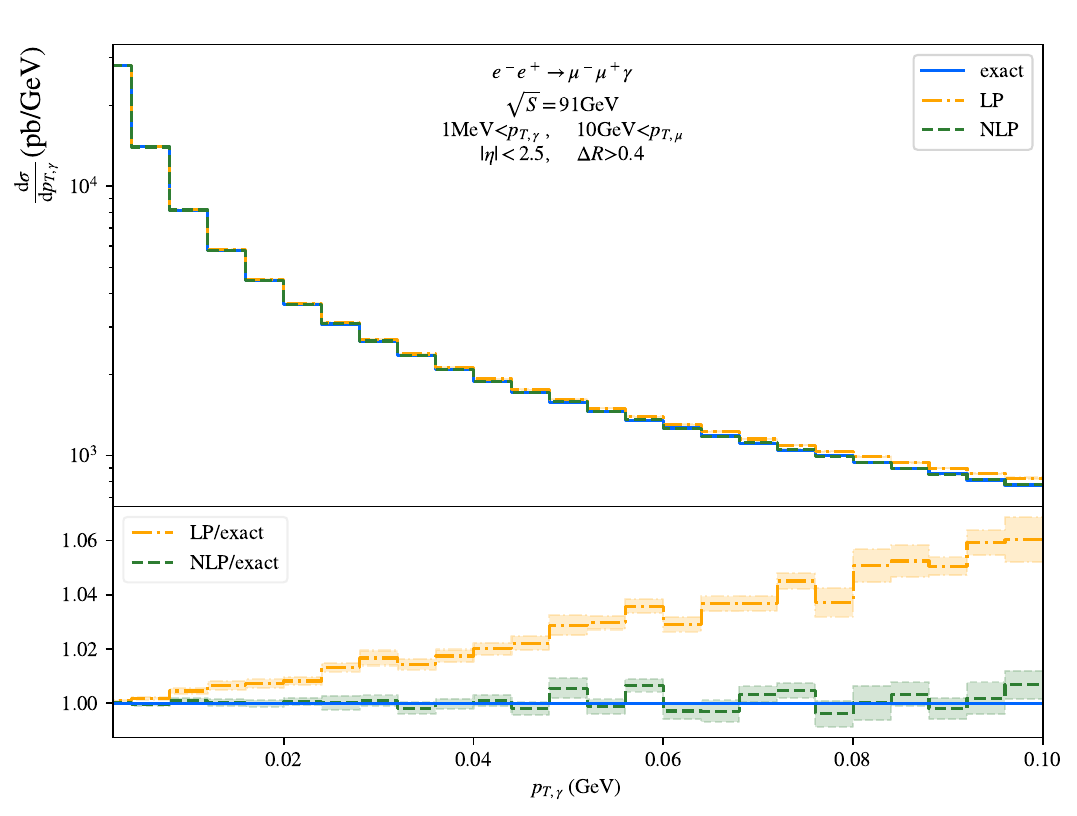}
    \end{subfigure}
    \begin{subfigure}[b]{0.495\textwidth}
        \includegraphics[width=\textwidth]{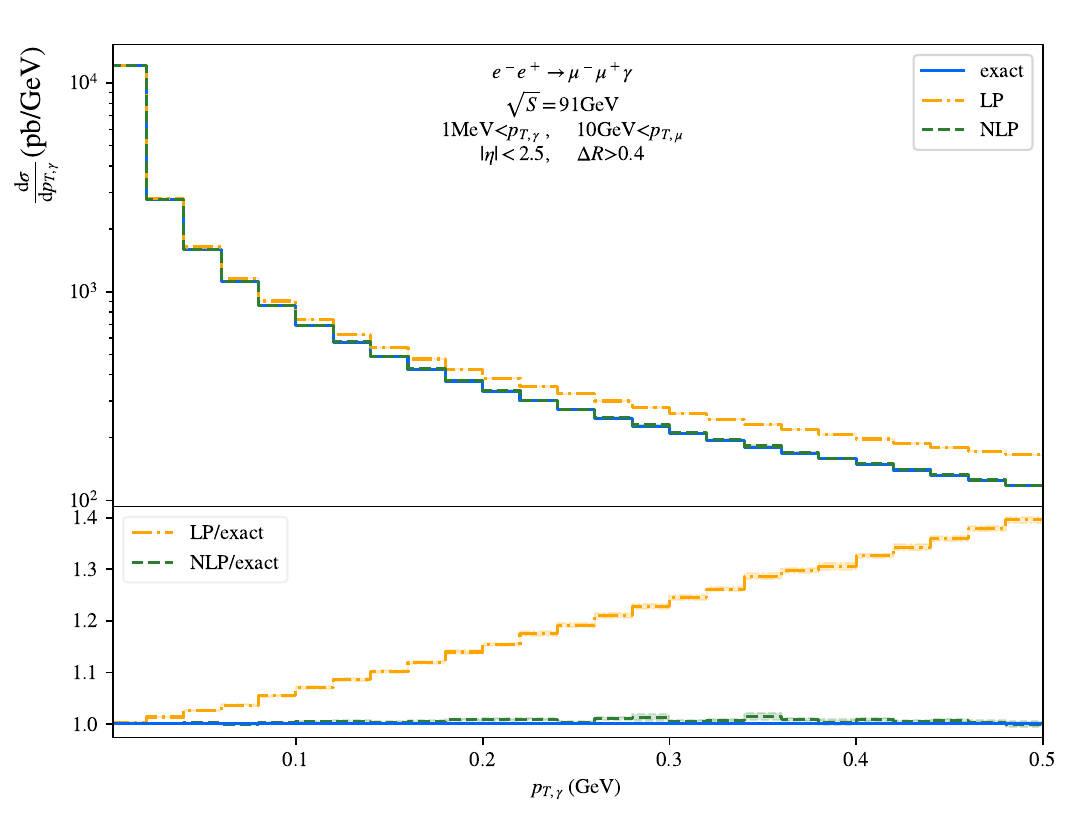}
    \end{subfigure}
    \\
    \begin{subfigure}[b]{0.495\textwidth}
        \includegraphics[width=\textwidth]{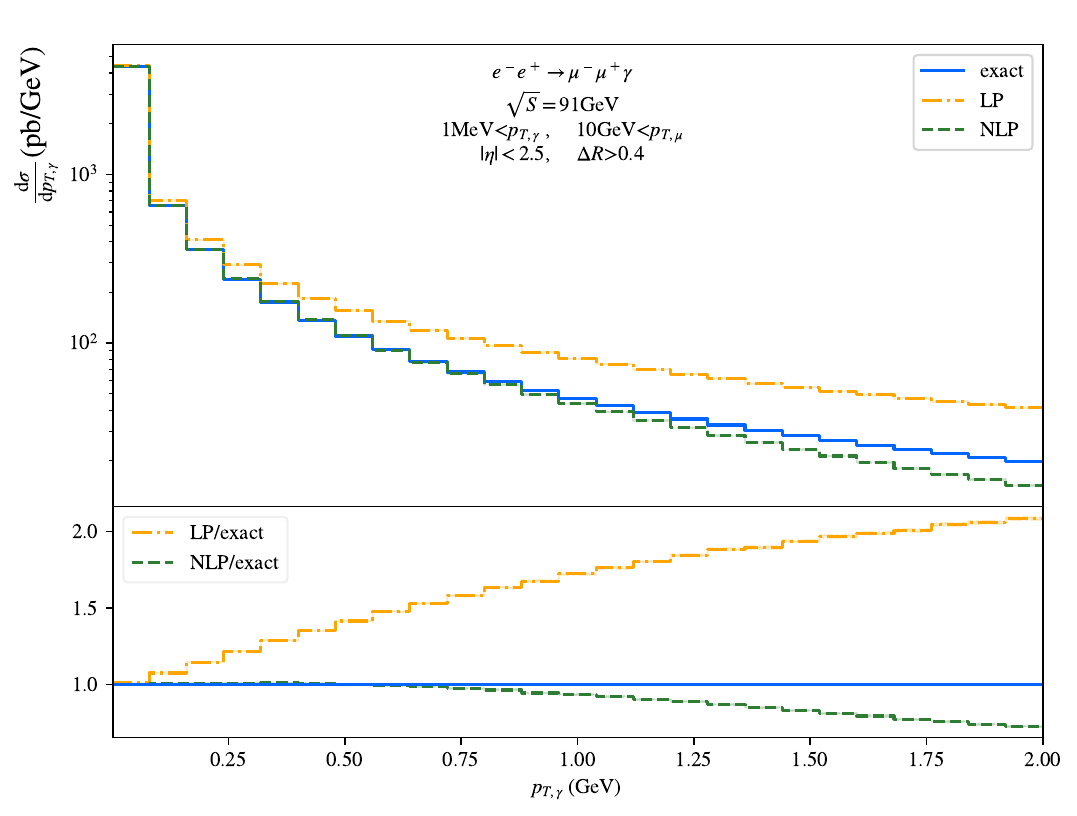}
        \label{NLP_ee_1MeV}
    \end{subfigure}
    \caption{The $\LP$ and $\NLP$ approximations and the exact result for the
    transverse momentum spectrum of the photon in the $e^- e^+
        \to \mu^-
        \mu^+ \gamma$ process at $\sqrt s =91$ GeV.}
    \label{fig:epempt}
\end{figure}

The NLP distributions in photon energy $\omega_\gamma$ and transverse
    momentum $p_{T, \gamma}$, obtained using the radiative amplitude with
    on-shell
    momenta \cref{modshifts}, are then compared to the LP and exact predictions
    in \cref{fig:epemomega} and \cref{fig:epempt}, respectively. In particular,
    we show distributions for very soft photons with 1 MeV$<  \omega_\gamma, \,
    p_{T, \gamma}< 100$ MeV. As
    discussed above, the NLP formula relying on shifting momenta on-shell
    returns predictions which provide a very good approximation of the exact
    result. Up to the scale of 100 MeV, the difference between the two
    predictions is at a few per mille level and grows to a 1-2\% level
    for $\omega_\gamma$ or $p_{T,\gamma}$ of up to ca. 1 GeV. In contrast, the
    LP approximation
    differs from the exact result by up to ca. 2\%  (6\%) and up to ca. 40\%
    (70\%) in these
    two ranges of $\omega_\gamma$ ($p_{T, \gamma}$), correspondingly.

\begin{figure}[htb]
    \centering
    \begin{subfigure}[b]{0.495\textwidth}
        \includegraphics[width=\textwidth]{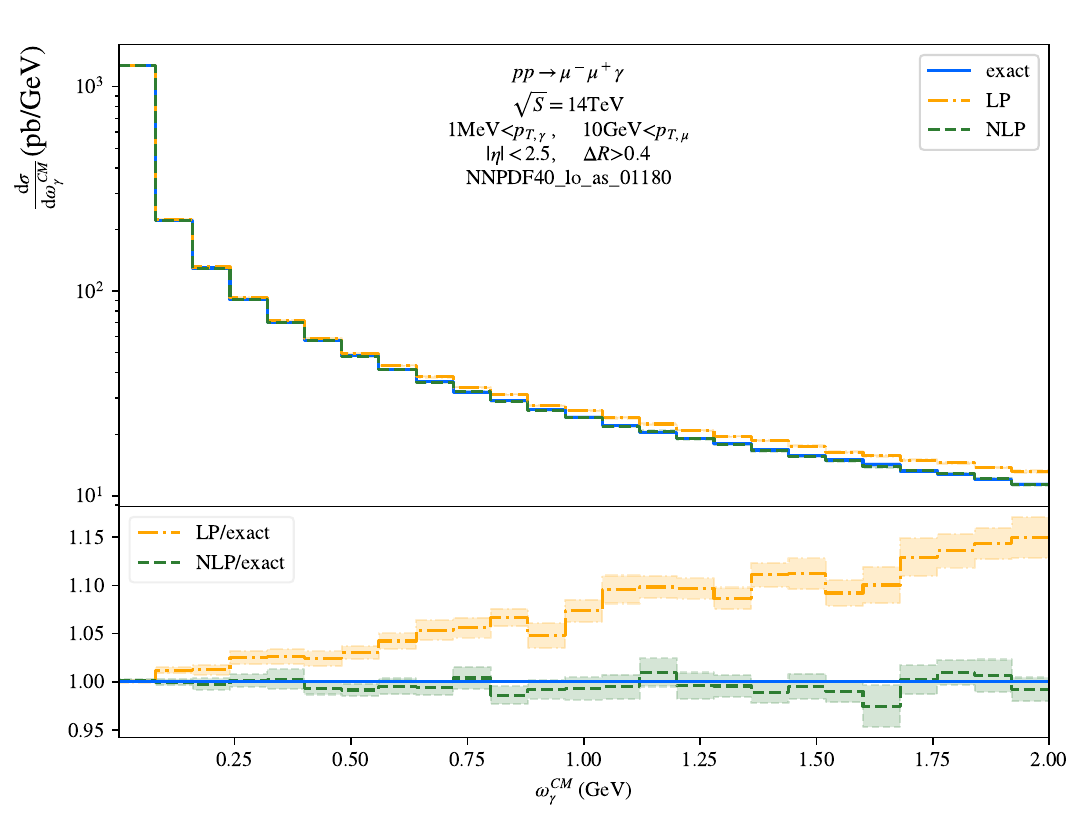}
    \end{subfigure}
    \begin{subfigure}[b]{0.495\textwidth}
        \includegraphics[width=\textwidth]{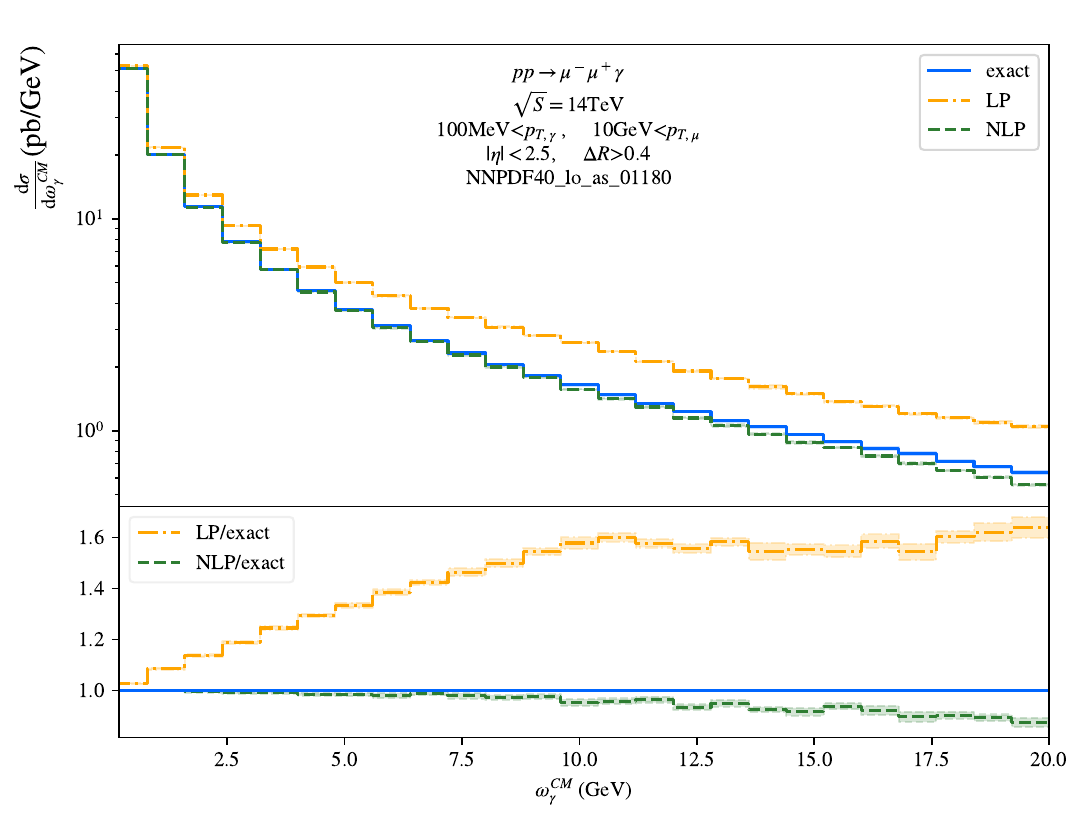}
    \end{subfigure}
    \\
    \begin{subfigure}[b]{0.495\textwidth}
        \includegraphics[width=\textwidth]{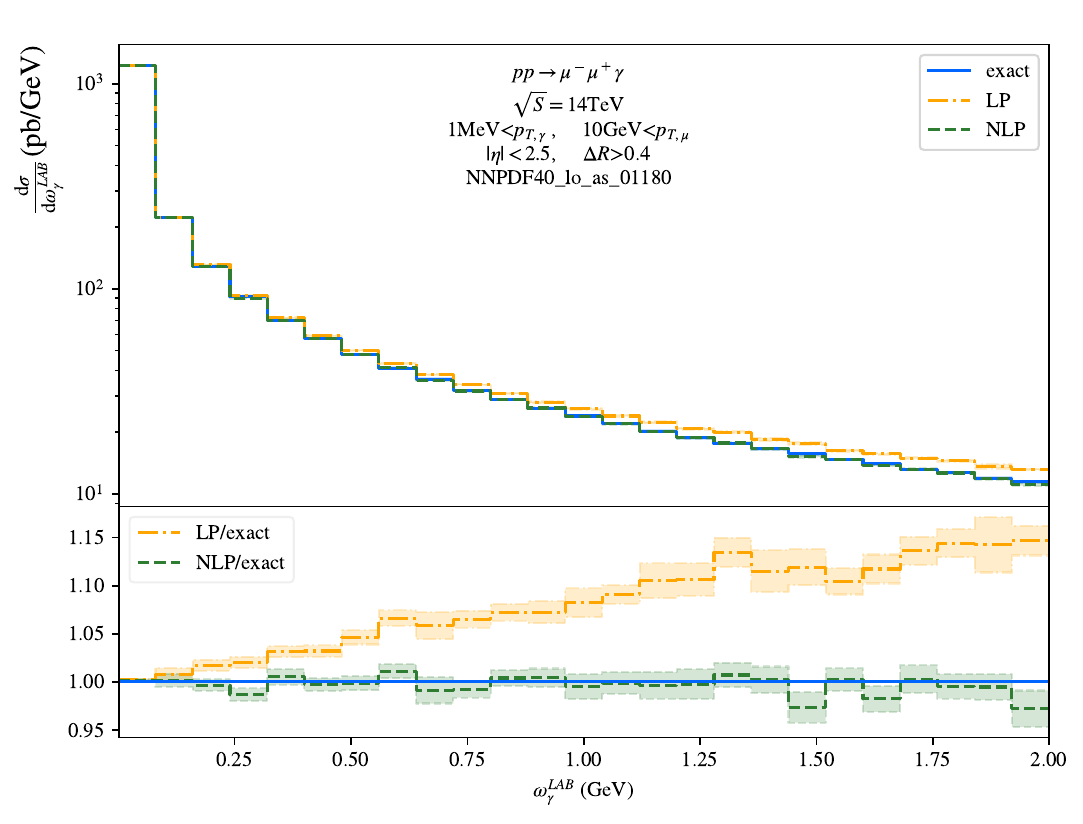}
    \end{subfigure}
    \begin{subfigure}[b]{0.495\textwidth}
        \includegraphics[width=\textwidth]{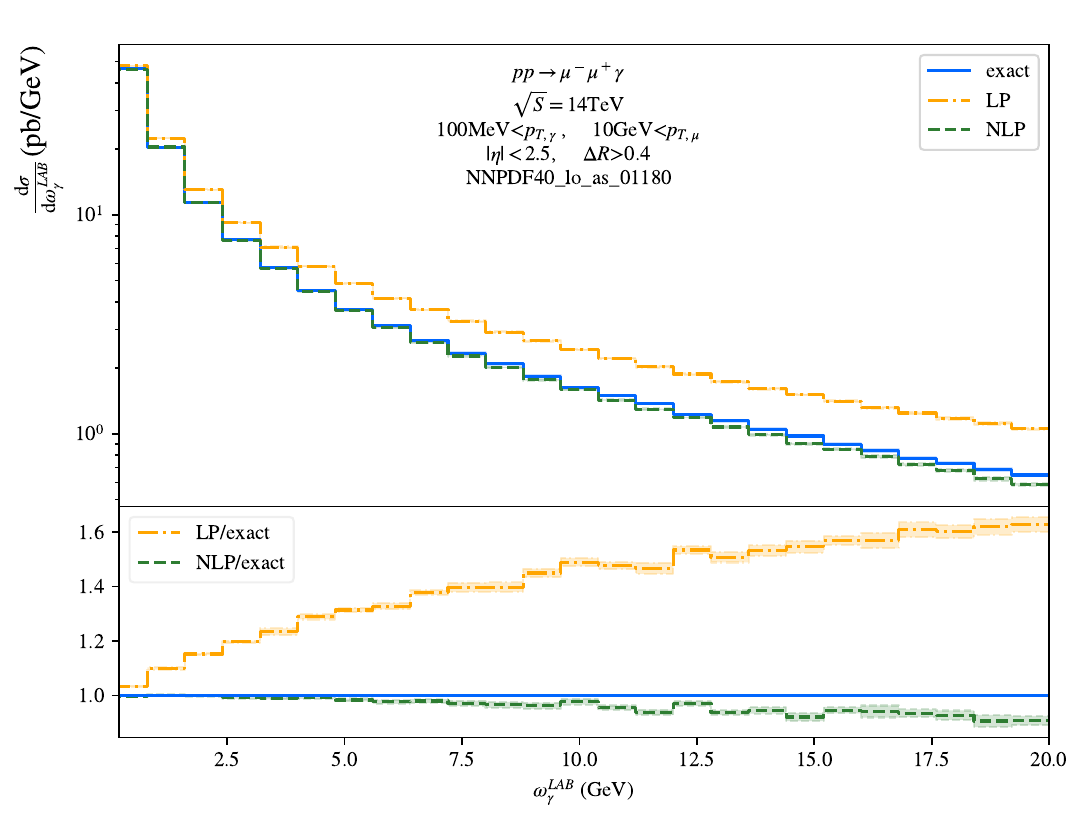}
    \end{subfigure}
        \caption{The $\LP$ and $\NLP$ approximations and the exact result for
        the energy spectrum of the photon in the partonic c.m. frame (top row)
        and the laboratory frame (bottom row) in the $pp
        \to \mu^-
        \mu^+ \gamma$ process at $\sqrt s =14$ TeV.}
    \label{fig:ppomega}
\end{figure}

\begin{figure}[htb]
    \centering
    \begin{subfigure}[b]{0.495\textwidth}
        \includegraphics[width=\textwidth]{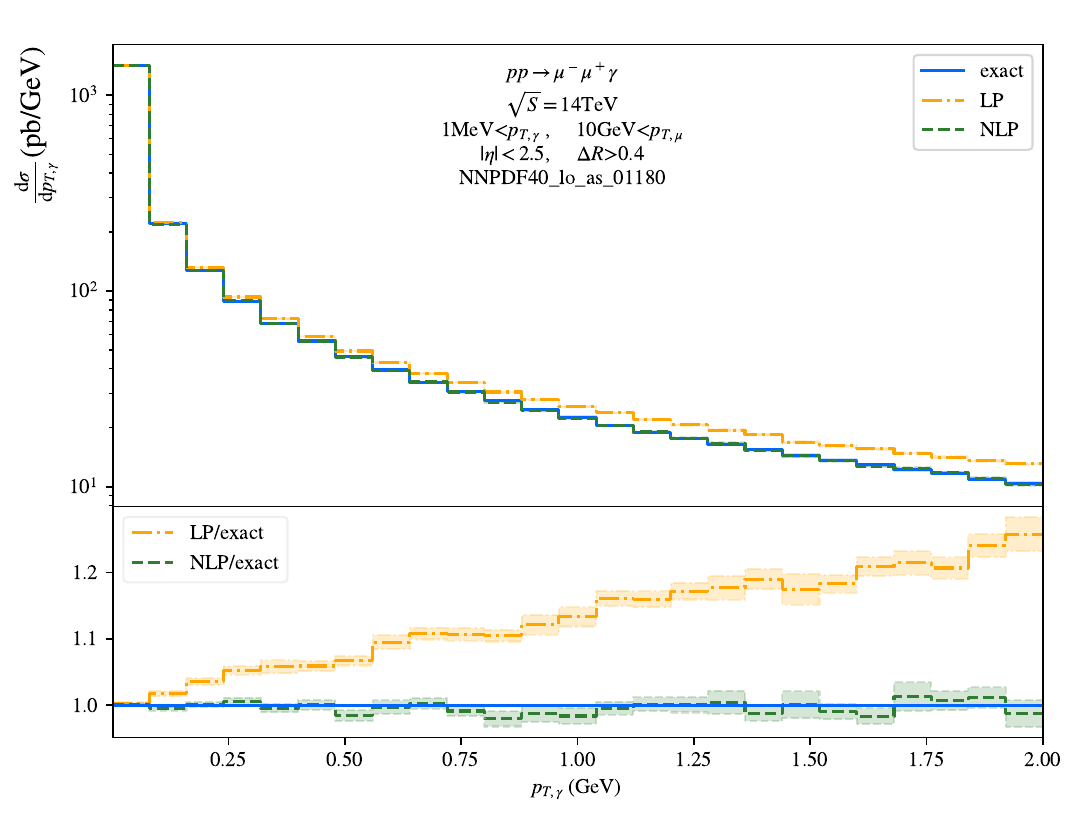}
    \end{subfigure}
    \begin{subfigure}[b]{0.495\textwidth}
        \includegraphics[width=\textwidth]{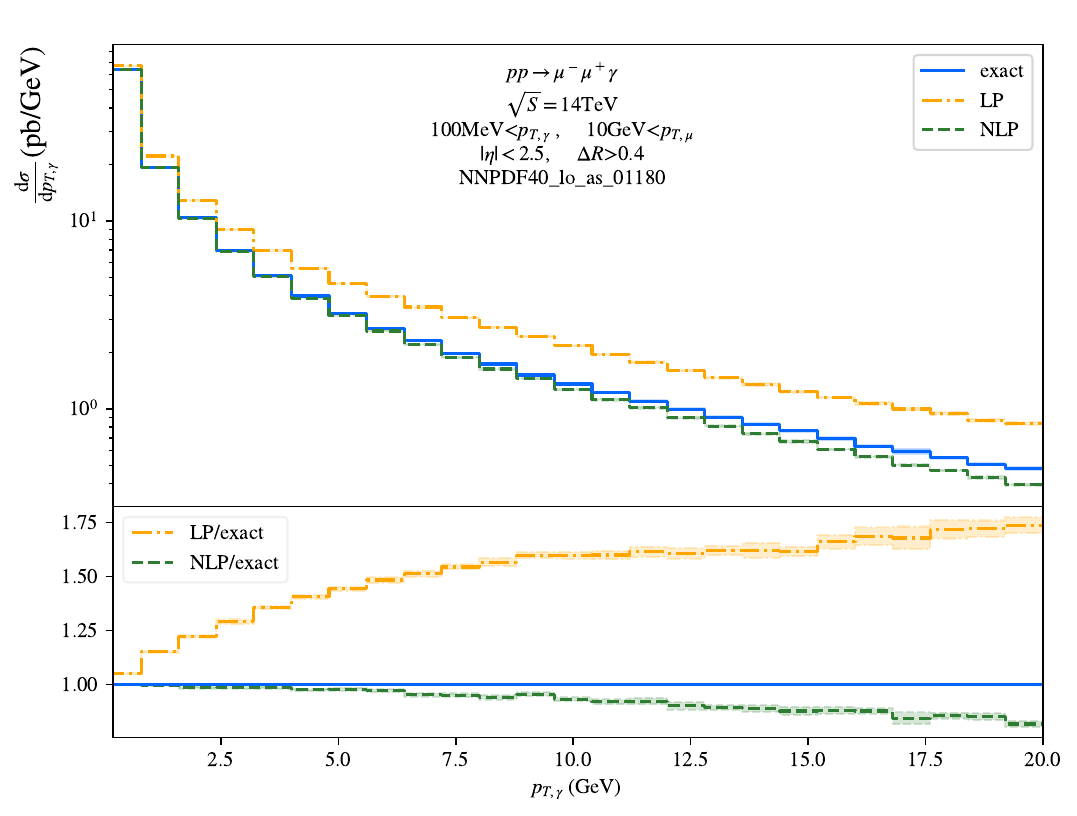}
    \end{subfigure}
        \caption{The $\LP$ and $\NLP$ approximations and the exact result for
        the transverse momentum spectrum of the photon  in the $pp
        \to \mu^-
        \mu^+ \gamma$ process at $\sqrt s =14$ TeV.}
    \label{fig:pppt}
\end{figure}

Next, we study the soft-photon spectra in the process $pp\to
\mu^+\mu^-\gamma$. The differential distributions in photon energy and
transverse momentum are shown in \cref{fig:ppomega} and \cref{fig:pppt},
respectively. The photon energy spectrum in \cref{fig:ppomega} is presented in
both the partonic c.m. frame and the laboratory frame. No perceptible
difference is observed between the results in the two frames for our choice of
kinematical cuts. Perhaps not
surprisingly, the behaviour of the LP and NLP approximations is very similar
to the one found for the $e^+e^-$ collisions. Quantitatively, however, in the
ranges of $\omega_\gamma$ and $p_{T, \gamma}$ studied here, the LP and NLP
predictions  appear to
be relatively closer to the exact results than in the $e^+e^-$ case.  To be
more precise, within an accuracy of roughly $10\%$, the LP spectrum deviates
from the exact result for $\omega_\gamma \gtrsim 400$MeV, $p_{T,\gamma}\gtrsim
150$MeV ($e^+e^-$) and   $\omega_\gamma \gtrsim
1$GeV, $p_{T, \gamma}\gtrsim 500$MeV ($pp$). The NLP predictions reach this
level of deviation only at $ p_{T, \gamma}\gtrsim 1$ GeV in the $e^+e^-$ case,
and at $ p_{T, \gamma}\gtrsim 10$ GeV  in the $pp$ case, i.e. outside of the
soft regime.


\section{Conclusions}
\label{sec:concl}
In this paper,  we have presented an extensive study of the LBK theorem and its
implications on numerical predictions for the soft-photon spectra.

 First, motivated by recent discussions in the
 literature\cite{Lebiedowicz:2021byo, Lebiedowicz:2023mlz,
 Lebiedowicz:2023ell}, we have addressed the consistency of the
 LBK theorem. As known for a long time, in the original formulation of the
 theorem\cite{Low:1958sn} the non-radiative amplitude is calculated on a set
 of unphysical momenta.  Seen in the most general way, this violation of
 momentum conservation leads to an ambiguity in the functional form of the
 non-radiative amplitude. We have provided a  proof that the aforementioned
 ambiguity can only affect the expansion of the radiative amplitude starting
 from NNLP in soft-photon energy, i.e. beyond the formal accuracy of the LBK
 theorem, thus proving the validity of the LBK theorem at NLP.
 In doing so, we have generalized the remark by Burnett and
 Kroll by observing an invariance of the theorem at NLP under a specific
 transformation of the non-radiative amplitude. The consequence of
this invariance is the presence of many equivalent forms of the LBK theorem,
 which include the original formulation by Low as one possibility.

Among the different versions of the theorem, for practical reasons, it is
particularly
attractive to consider those that restore  momentum conservation in the
calculation of the non-radiative amplitude.  Such restoration can be achieved
by reformulating the LBK theorem in terms of the non-radiative amplitude
calculated on momenta which values are modified wrt. the momenta of the
radiative amplitude. The modification that has been put forward in the
literature \cite{DelDuca:2017twk, vanBeekveld:2019prq, Bonocore:2021cbv} relies
on adding small shifts of the
order of the soft-photon momentum
$k$. Apart from reviewing the derivation of the reformulated LBK theorem in
terms of shifted momenta, we have proposed expressions for the momenta shifts
which not only restore momentum conservation,  but also ensure that each of the
shifted momenta is on-shell. In this way, we facilitate the generation and
numerical calculation of the non-radiative amplitude with a wide range of
publicly available tools.

We have also studied the quality of the soft approximation
provided by the LBK theorem in its various formulations on
the example of the
$e^+e^-  \to \mu^+ \mu^- \gamma$ process, analysed at LEP1 energies.  We have
found a remarkable improvement in the quality of the approximation for the
formulations involving shifts w.r.t. the formulation of the theorem with
derivatives, although we have considered the latter only for a specific form of
the
non-radiative
amplitude. Depending on the required quality of the approximation, NNLP effects
can be thus numerically relevant and the form of the LBK theorem used to
calculate the NLP approximation of the soft-photon spectra needs to be chosen
carefully. In this regard, our studies indicate that the formulations involving
shifts should be preferred. Notably, for the implementation involving on-shell
shifts, we have used the amplitudes generated by the MadGraph5 code, in this
way
demonstrating the feasibility of the calculations for a wide range of
processes,
given a corresponding phase-space integrating code is available.

Despite the long history of the LBK theorem, to the best of our knowledge, no
study in the literature has explicitly identified when
power-suppressed effects become visible in the soft-photon spectra. In order to
address this question, we compared the exact, LP and NLP results for the
$e^+e^-  \to \mu^+ \mu^- \gamma$ process at the LEP1 energy of $91$ GeV and the
$pp \to \mu^+
\mu^- \gamma$ process at the LHC energy of $14$ TeV. We have studied various
ranges of photon energy and transverse momentum. For the cases
studied here,  the quality of the NLP approximation  in a reasonably soft
regime
is of the order of, or better than, one percent, even though the specific
results
depend on the process, observable and the analysis set-up. The quality of the
LP
approximation is significantly worse, meaning that for measurements where a
precision of a few percent can be reached in the soft regime, the LP might not
provide a good enough approximation. Correspondingly, our results suggest that
in order to access the power-suppressed terms, a percent-level precision is
needed, especially in the case of the $pp \to \mu^+ \mu^- \gamma$ process at
the
LHC. Obviously, this is only a  crude estimate. A more precise
statement would require a careful analysis of all  theoretical and experimental
uncertainties.

The results of this work open up the possibility of many follow-up analyses.
The  two processes we studied here involve simple leptonic final states and
central rapidity photons. Given that the excess in the soft-photon spectrum was
observed for $e^+e^-$ collisions in hadronic final states, it would be
interesting to investigate the impact of NLP corrections on the predictions for
the associated photon production with jets. In this regard, it would also be
interesting to extend the analysis by including QCD corrections at
1-loop\cite{Bonocore:2021cbv}, which will be relevant for both initial and
final state hadrons. In the long run, the studies need to be extended to $pp$
collisions resulting in various hadronic final states with photons at forward
rapidities, as planned to be investigated with the ALICE 3
detector\cite{ALICE:2022wwr}.

\section*{Acknowledgments}
We would like to thank A. Andronic, V. Hirschi, F. Kling and S. Weinzierl for
useful discussions. We would also like to thank the MadGraph, Sherpa and Vegas+
authors for helping us with their software.
DB would like to thank the
participants to the EMMI workshop
``Real and virtual photon production at ultra-low transverse momentum and
low mass at LHC'' and in particular P. Braun-Munziger, X. Feal, S.
Fl\"orchinger, H. van Hees, P.
Lebiedowicz, O. Nachtmann, K. Reygers, K. Schweda, J. Stachel and M. V\"olkl for
stimulating discussions on the topic of soft photons.
DB and AK are grateful to
the Galileo Galilei
Institute for hospitality and
support during the
scientific program ``Theory Challenges in the Precision Era of the Large
Hadron
Collider'', where part of this work was done. AK thanks the CERN Theory
Division
for the hospitality and support during the early stages of this work.
The work of RB was partially supported by the Deutsche Forschungsgemeinschaft
(DFG) through
the
Research Training Group ``GRK 2149: Strong and Weak Interactions: from Hadrons
to Dark Matter''.   The work of DB was supported by the Excellence Cluster
ORIGINS funded by the DFG under
Grant No. EXC- 2094-390783311.

\appendix
\section{Case study: $2\to2$ scattering}
\label{sec:comp}

In this section we apply the general arguments of \cref{sec:validity}
concerning the traditional form of the LBK theorem with derivatives
to the simple case of photon bremsstrahlung in pion scattering. In doing so,
we compare with the analysis of Lebiedowicz, Nachmann, Szczurek (LNS)
\cite{Lebiedowicz:2021byo}, showing that their result is equivalent to the
traditional form of the LBK
theorem and with previous studies in the
literature.

\subsection{LBK theorem for $\pi^-\pi^0 \to \pi^-\pi^0\gamma$}
Following  \cite{Low:1958sn} and \cite{Lebiedowicz:2021byo},
and using the notation established in \cite{Lebiedowicz:2021byo},
we consider the following process:
$$\pi^-(p_a)\pi^0(p_b) \to \pi^-(p'_1)\pi^0(p'_2) \gamma(k)~,$$
with $$p_a + p_b = p'_1 + p'_2 + k~.$$
For this process, the LBK theorem in the form of \cref{ext+int}, adapted to the
case of scalar particles,
reads
\begin{align}
\mathcal{A}^\mu(p_a, p_b, p'_1, p'_2, k)
&= e\left[
\frac{p_a^\mu}{k\cdot p_a} \mathcal{H}(p_a, p_b, p'_1, p'_2)
+G_a^{\mu \nu}\pdv{\mathcal{H}(p_a, p_b, p'_1, p'_2)}{p_{a}^\nu}
\right]\notag\\&\phantom{=}
- e\left[
\frac{p_1^{\prime\mu}}{k\cdot p'_1} \mathcal{H}(p_a, p_b, p'_1, p'_2)
- G_1^{\mu \nu}\pdv{\mathcal{H}(p_a, p_b, p'_1, p'_2)}{p_{1}^{\prime\nu}}
\right] + \order{k}\notag\\&
= e\left[\frac{p_a^\mu}{k\cdot p_a} - \frac{p_1^{\prime\mu}}{k\cdot p'_1}
+G_a^{\mu \nu} \pdv{p_a^\nu} + G_1^{\mu \nu}\pdv{p^{\prime\nu}_1}\right]
\mathcal{H}(p_a, p_b, p'_1, p'_2) + \order{k}~.
\label{A1}
\end{align}
To show the ambiguity in the calculation of the non-radiative amplitude with
radiative kinematics, we consider the following two choices for $\mathcal{H}$:
\begin{equation}
\mathcal{H}_1 = \mathcal{H}(s'_L, t_1), \qquad \mathcal{H}_2 =
\mathcal{H}(s'_L, t_2)~,
\end{equation}
where we defined
\begin{equation}\label{sL}
s'_L = p_a\cdot p_b + p'_1\cdot p'_2, \qquad
t_1 = (p_a - p'_1)^2, \qquad t_2 = (p_b - p'_2)^2.
\end{equation}
Here, $\mathcal{H}(s_L, t)$ is the amplitude for the non-radiative process
$\pi^-(p_a)\pi^0(p_b) \to \pi^-(p_1)\pi^0(p_2)$.
Obviously, $p_a + p_b = p_1 + p_2$, so in the elastic limit
$t_1 = (p_a - p_1)^2$ and $t_2 = (p_b - p_2)^2$, hence we have $t_1 = t_2$,
and thus $\mathcal{H}_1=\mathcal{H}_2$.

Now we consider
\begin{align}
\pdv{\mathcal{H}_i}{p_a^\nu} &= \pdv{\mathcal{H}_i}{s'_L}\pdv{s'_L}{p_a^\nu} +
\pdv{\mathcal{H}_i}{t_i}\pdv{t_i}{p_a^\nu}, \notag \\
\pdv{\mathcal{H}_i}{p^{\prime\nu}_1} &=
\pdv{\mathcal{H}_i}{s'_L}\pdv{s'_L}{p^{\prime\nu}_1} +
\pdv{\mathcal{H}_i}{t_i}\pdv{t_i}{p^{\prime\nu}_1}~,
\label{A4}
\end{align}
where no sum over the index $i=1,2$ has been assumed.
The derivatives of $s'_L$ in \cref{A4} read
\begin{equation*}
\pdv{s'_L}{p_a^\nu} = p_{b\nu}, \qquad \pdv{s'_L}{p^{\prime\nu}_1} = p'_{2\nu}~.
\end{equation*}
The terms with derivatives of $t_i$ are
different for $i=1,2$, and are given by
\begin{equation}
\pdv{t_1}{p^\nu_a} = 2(p_a-p'_1)_\nu, \qquad \pdv{t_1}{p^{\prime\nu}_1} =
-2(p_a-p'_1)_\nu, \qquad \pdv{t_2}{p^\nu_a}=\pdv{t_2}{p^{\prime\nu}_1}=0~.
\end{equation}
Putting the above equations together we obtain two seemingly distinct forms for
the LBK theorem, depending on whether we take ${\cal H}={\cal H}_1$
or ${\cal H}={\cal H}_2$ in \cref{A1}. They read
\begin{align}
\mathcal{A}^\mu(p_a, p_b, p'_1, p'_2, k)
= e\Bigg[&\left(\frac{p_a^\mu}{k\cdot p_a} - \frac{p_1^{\prime\mu}}{k\cdot
    p'_1}\right) + \left(p_b^\mu-\frac{p_b \cdot k}{p_a\cdot k}p_a^\mu +
p^{\prime\mu}_2-\frac{p'_2\cdot k}{p'_1\cdot
    k}p^{\prime\mu}_1\right)\pdv{s'_L}\notag\\& -2(p_a-p'_1)\cdot
k\left(\frac{p_a^\mu}{k\cdot p_a} - \frac{p_1^{\prime\mu}}{k\cdot
    p'_1}\right)\pdv{t_1}\Bigg] \mathcal{H}_1 + \order{k}~,\notag \\
\mathcal{A}^\mu(p_a, p_b, p'_1, p'_2, k)
= e\Bigg[&\left(\frac{p_a^\mu}{k\cdot p_a} - \frac{p_1^{\prime\mu}}{k\cdot
    p'_1}\right)\notag \\
& + \left(p_b^\mu-\frac{p_b \cdot k}{p_a\cdot k}p_a^\mu +
p^{\prime\mu}_2-\frac{p'_2\cdot k}{p'_1\cdot
    k}p^{\prime\mu}_1\right)\pdv{s'_L}\Bigg] \mathcal{H}_2 + \order{k}~.
\label{LBKpion}
\end{align}
If one ignores the fact that $\mathcal{H}_1$ and $\mathcal{H}_2$ are not the
same, the two expressions indeed yield different results, thus seemingly
invalidating
the LBK theorem. However, by relating $\mathcal{H}_1$ and $\mathcal{H}_2$, the
difference disappears at NLP. In fact, one has
\begin{equation}
    \mathcal{H}_2
    = \mathcal{H}(s'_L, t_2)
    = \mathcal{H}(s'_L, t_1)
    + (t_2 - t_1) \pdv{\mathcal{H}(s'_L, t_1)}{t_1}
    + \order{k^2}~.
\end{equation}
Using then the expression
\begin{equation}
t_2-t_1 = (p_b - p'_2)^2 - (p_a - p'_1)^2 = (p_a - p'_1 - k)^2 - (p_a - p'_1)^2
= -2k\cdot (p_a - p'_1) + \order{k^2}~,
\end{equation}
we get
\begin{equation}
    \mathcal{H}_2
    = \mathcal{H}_1 - 2k \cdot (p_a - p'_1) \pdv{\mathcal{H}_1}{t_1}
    + \order{k^2}~,
\end{equation}
which accounts for the difference between the two previous expressions.
This is the cancellation we saw in \cref{sec:validity},
where we proved it in the general case.

\subsection{Comparison between LNS and the original work of Low}

    The soft bremsstrahlung in pion scattering has been computed also by the
    authors of ref.
 \cite{Lebiedowicz:2021byo}. Their final result, which is given by
eq. (3.27) of their paper, reads
\begin{align}\label{LNS}
\mathcal{A}^\mu = &e \mathcal{H}(s_L, t) \left[\frac{p_a^\mu}{p_a\cdot k} -
\frac{p^{\prime\mu}_1}{p'_1\cdot k}\right] + 2e \pdv{\mathcal{H}(s_L,
t)}{s_L}\left[p_b^\mu - \frac{p_b\cdot k}{p_a\cdot k}p_a^\mu\right] \nonumber\\&
-2e\pdv{\mathcal{H}(s_L, t)}{t}\left[(p_a-p_1)\cdot k - p_a\cdot
l_1\right]\left[\frac{p_a^\mu}{p_a\cdot k} - \frac{p^{\mu}_1}{p_1\cdot
k}\right] + \order{k}~,
\end{align}
where the non-radiative amplitude $\cal H$ is written as a function of the two
variables $s_L$
and $t$
\begin{equation}
s_L = p_a\cdot p_b + p_1\cdot p_2, \qquad t=(p_a-p_1)^2=(p_b-p_2)^2.
\end{equation}
The momenta fulfill the relations
\begin{equation}
p_a + p_b = p_1 + p_2 = p'_1 + p'_2 + k
\end{equation}
and $l_i$ are defined as a shift between $p$ and $p'$ as follows:
\begin{equation}
l_i = p_i - p'_i.
\end{equation}
As done in \cite{Low:1958sn}, here we set $k^2=0$ and drop
all terms proportional to $k^\mu$ since we assume
all final states
to be on-shell.

The authors of \cite{Lebiedowicz:2021byo} compare then \cref{LNS} to the
one
derived by Low in ref. \cite{Low:1958sn}, which they report in eq. (3.29) of
their paper. It reads
\begin{equation}\label{wLow}
\tilde{\mathcal{A}}^\mu = e \mathcal{H}(s_L, t) \left[\frac{p_a^\mu}{p_a\cdot
k} - \frac{p^{\mu}_1}{p_1\cdot k}\right] + e \pdv{\mathcal{H}(s_L,
t)}{s_L}\left[p_b^\mu - \frac{p_b\cdot k}{p_a\cdot k}p_a^\mu + p_2^\mu -
\frac{p_2\cdot k}{p_1 \cdot k}p_1^\mu\right] + \order{k}~.
\end{equation}
They conclude that, while the LP terms
agree, there is a discrepancy in the NLP terms.
We first point out that \eqref{wLow} does not precisely
coincide with the result given by Low in equation (2.16) of \cite{Low:1958sn}.
In fact, by looking at eq. (2.1) and eq. (2.16) in \cite{Low:1958sn}, one
concludes that the correct
expression should be\footnote{In the final stages of writing this paper, we
have been informed that the authors are aware of this mistake. }
\begin{equation}\label{Low}
\mathcal{A}^\mu = e \mathcal{H}(s'_L, t_2) \left[\frac{p_a^\mu}{p_a\cdot k} -
\frac{p^{\prime\mu}_1}{p'_1\cdot k}\right] + e \pdv{\mathcal{H}(s'_L,
t_2)}{s'_L}\left[p_b^\mu - \frac{p_b\cdot k}{p_a\cdot k}p_a^\mu +
p^{\prime\mu}_2 - \frac{p'_2\cdot k}{p'_1 \cdot k}p^{\prime\mu}_1\right] +
\order{k}~,
\end{equation}
with $s'_L$ and $t_2$ defined as in \cref{sL}.
This form of the LBK theorem is indeed what we obtained in \cref{LBKpion}.

A careful comparison of expressions \cref{LNS} and \cref{Low}
shows that they are in perfect agreement with each other, up to $\order{k}$. To
simplify the comparison, it is worth noticing that $p$ and
$p'$ are equal up to $\order{k}$ corrections. Therefore, to order $\order{k}$,
the
difference between $p$ and $p'$ is only relevant for the first term, and
\cref{Low} can be rewritten as
\begin{equation}\label{E7}
\mathcal{A}^\mu = e \mathcal{H}(s'_L, t_2) \left[\frac{p_a^\mu}{p_a\cdot k} -
\frac{p^{\prime\mu}_1}{p'_1\cdot k}\right] + e \pdv{\mathcal{H}(s_L,
t)}{s_L}\left[p_b^\mu - \frac{p_b\cdot k}{p_a\cdot k}p_a^\mu + p^\mu_2 -
\frac{p_2\cdot k}{p_1 \cdot k}p^\mu_1\right] + \order{k}~.
\end{equation}
To compare \cref{LNS} with \cref{E7} it is necessary to use the formula
\begin{equation}\label{expansion}
\mathcal{H}(s'_L, t_2) = \mathcal{H}(s_L, t) + \delta s'_L
\pdv{\mathcal{H}}{s_L} + \delta t_2 \pdv{\mathcal{H}}{t}~,
\end{equation}
where we defined
\begin{align}
    \delta s'_L &
    = s'_L - s_L
    = p'_1 \cdot p'_2 - p_1 \cdot p_2
    = (p_1 - l_1) \cdot (p_2 - l_2) - p_1 \cdot p_2 \nonumber\\&
    = -(l_1 \cdot p_2 + l_2 \cdot p_1) + \order{k^2}
    = -(p_1 + p_2) \cdot k + \order{k^2} \nonumber\\&
    = -(p_a + p_b) \cdot k + \order{k^2}~, \notag \\
    \delta t_2 &
    = t_2 - t = (p_b - p'_2)^2 - (p_b - p_2)^2
    = (p_b - p_2 + l_2)^2 - (p_b - p_2)^2 \nonumber \\&
    = 2l_2 \cdot (p_b - p_2) + \order{k^2}
    = -2(k - l_1) \cdot (p_a - p_1) + \order{k^2} \nonumber \\&
    = -2\left[(p_a - p_1) \cdot k - p_a \cdot l_1\right] + \order{k^2}~.
\end{align}
Here we used the relations $p_1\cdot l_2 = p_1\cdot k$ and $p_2\cdot l_1 =
p_2\cdot k$, which can be derived from the two relations $l_1+l_2=k$ and
$p_i\cdot l_i = 0$ (i.e. eq. (3.17) and eq. (3.22) in
\cite{Lebiedowicz:2021byo}).

Inserting now  \cref{expansion} into \cref{E7}, we see that the term
proportional to $\mathcal{H}$ is identical to the one in \cref{LNS}.
Additionally, the
term proportional to $\pdv{\mathcal{H}}{t}$  is given by
\begin{equation}
e \delta t_2 \pdv{\mathcal{H}}{t} \left[\frac{p_a^\mu}{p_a\cdot k} -
\frac{p^{\prime\mu}_1}{p'_1\cdot k}\right]
= -2e \pdv{\mathcal{H}}{t}\left[(p_a-p_1)\cdot k - p_a\cdot l_1 \right]
\left[\frac{p_a^\mu}{p_a\cdot k} - \frac{p^{\prime\mu}_1}{p'_1\cdot k}\right]~.
\label{A19}
\end{equation}
The r.h.s. of \cref{A19}  coincides with the third term in \cref{LNS}
after
dropping
the prime
in the last parenthesis (this is possible because this term is already a NLP
term, and thus the difference due to replacing $p$ with $p'$ is a NNLP effect).
Finally, the
term proportional to $\pdv{\mathcal{H}}{s_L}$ has now the following expression
\begin{equation}
e \delta s'_L \pdv{\mathcal{H}}{s_L} \left[\frac{p_a^\mu}{p_a\cdot k} -
\frac{p^{\mu}_1}{p_1\cdot k}\right] + e \pdv{\mathcal{H}}{s_L}\left[p_b^\mu -
\frac{p_b\cdot k}{p_a\cdot k}p_a^\mu + p^\mu_2 - \frac{p_2\cdot k}{p_1 \cdot
k}p^\mu_1\right]~,
\end{equation}
where again the prime in the first term can dropped because $\delta s'_L$ is
already of order $\order{k}$. After some algebra, it is easy to show that this
term reads
\begin{align}
2e \pdv{\mathcal{H}}{s_L}\left[p_b^\mu - k\cdot p_b\frac{p_a^\mu}{p_a\cdot
k}\right]~,
\end{align}
in perfect agreement with the second term in
\cref{LNS}. Therefore, the LNS result (\cref{LNS}) and Low's
original result (\cref{Low})
are completely equivalent at NLP.

We note also that although here we focused on pion scattering, we expect
analogous arguments to hold for the proton scattering discussed in
\cite{Lebiedowicz:2023ell, Lebiedowicz:2023mlz}.

\subsection{Comparison between LNS and other literature}

A comparison between the result of LNS, i.e. \cref{LNS}, and the previous
literature has been carried out in the appendix of
\cite{Lebiedowicz:2021byo}. It is claimed there that all previous known forms
of the LBK theorem are problematic. In particular, LNS claim that the result of
reference
\cite{Gervais:2017yxv} is not consistent, since it
depends on an arbitrary quantity, as discussed below. However, a careful
analysis
shows  that this claim has no valid foundation, since the dependence on
such quantity vanishes.

The argument in \cite{Lebiedowicz:2021byo} is the following.
In \cite{Gervais:2017yxv} Low's theorem is written in a form that depends on
the following four quantities:
\begin{align}
I_1 &= \left(-l_1 \cdot \pdv{p_1} - l_2 \cdot \pdv{p_2} - k \cdot
\pdv{p_a}\right)\mathcal{H}(p_a, p_b, p_1, p_2)~, \notag \\
I_2 &= \left(-l_1 \cdot \pdv{p_1} - l_2 \cdot \pdv{p_2} + k \cdot
\pdv{p_1}\right)\mathcal{H}(p_a, p_b, p_1, p_2)~, \notag \\
I_3^\mu &= \pdv{\mathcal{H}(p_a, p_b, p_1, p_2)}{p_{a\mu}}, \qquad I_4^\mu =
\pdv{\mathcal{H}(p_a, p_b, p_1, p_2)}{p_{1\mu}}~.\label{I}
\end{align}
If one then restricts the analysis to an amplitude $\mathcal{H}$ that depends
solely on the
quantity $p_a^2+p_1^2-p_b^2-p_2^2$, as done in \cite{Lebiedowicz:2021byo},
the elastic amplitude is given by a constant, since \footnote{
    Note that
    the definition of $\eta$ and $\xi$ given by eq.~(14) in
    \cite{Gervais:2017yxv} implies
    that the elastic momenta are not on-shell.
    However, this detail is irrelevant for our discussion in this section.
}
\begin{align}
\mathcal{H}(p_a,p_b,p_1,p_2) = f(p_a^2+p_1^2-p_b^2-p_2^2) = f(m_a^2 + m_1^2 -
m_b^2 - m_2^2) \equiv f_0.
\label{A24}
\end{align}
Thus, the derivatives of the elastic amplitude vanish and the final result only
depends on the value of $f_0$.
However, if one considers the corresponding expressions
in \cref{I}, they become
\begin{equation}
I_1 = -2(k\cdot p_a)f'_0, \qquad I_2 = 2(k\cdot p_1)f'_0, \qquad I_3^\mu =
2p_a^\mu f'_0, \qquad I_4^\mu = 2p_1^\mu f'_0~.
\label{f0}
\end{equation}
Therefore, in \cref{f0} there is a dependence on $f'_0$ (the derivative of $f$
evaluated on
non-radiative momenta). $f'_0$ is an arbitrary quantity, which
seems to
invalidate the consistency of the result in \cite{Gervais:2017yxv}.

To see why this argument does not imply an inconsistency of the LBK theorem in
the
form given in ref. \cite{Gervais:2017yxv}, a more detailed analysis of
the complete
expressions is needed.
A direct comparison between \cite{Gervais:2017yxv} and
\cite{Lebiedowicz:2021byo} is unfortunately not possible, since the processes
under consideration are
different ($e^- \pi^0 \to e^- \pi^0$ for ref. \cite{Gervais:2017yxv}, and $\pi^-
\pi^0 \to \pi^- \pi^0$ for LNS). The comparison thus requires a dictionary to
relate the
theorems with fermions and scalar fields, respectively, which is summarized in
Table \ref{T1}.
\begin{table}[]
    \centering
    \begin{tabular}{c|c}
        Gervais & Lebiedowicz, Nachtmann, Szczurek \\
        \hline
        $\frac{1}{\s{p}-m}$ & $\frac{1}{p^2-m^2}$ \\
        $\gamma_\mu$ & $p_\mu + p'_\mu$ \\
        $u, v, \bar{u}, \bar{v}$ & $1$ \\
        $p_1$, $p'_1$ & $p_a$ \\
        $k_1$, $k'_1$ & $p_b$ \\
        $p_2$ & $p'_1$ \\
        $k_2$ & $p'_2$ \\
        $p'_2$ & $p_1$ \\
        $k'_2$ & $p_2$ \\
        $q$ & $k$ \\
        $\eta_1$, $\xi_1$ & 0 \\
        $\xi_2$ & $-l_1$ \\
        $\eta_2$ & $-l_2$ \\
    \end{tabular}
    \caption{Relation between the notations used in \cite{Gervais:2017yxv} and
    \cite{Lebiedowicz:2021byo}.}
    \label{T1}
\end{table}
After carefully
taking this translation into account,
Low's theorem in the notation of ref. \cite{Gervais:2017yxv}
(see equation (20) there) reads
\begin{equation}
\mathcal{A}^\mu = e \left(\frac{p_1^{\prime\mu}}{p'_1\cdot k} -
\frac{p_a^\mu}{p_a\cdot k}\right)\mathcal{H}(p_a,p_b,p_1,p_2) - e
\left(\frac{p_a^\mu}{p_a\cdot k}I_1 - \frac{p_1^\mu}{p_1\cdot k}I_2 + I_3^\mu +
I_4^\mu\right)~.
\label{Gerv}
\end{equation}
In particular, it is worth noting that the expression for $I_i$ in \cref{Gerv}
enter via the
following combinations
\begin{equation}
p_a^\mu I_1 + (p_a\cdot k)I^\mu_3, \qquad p_1^\mu I_2 - (p_1\cdot k)I_4^\mu~.
\label{Ivanish}
\end{equation}
Therefore, after inserting the expressions of \cref{f0} in \cref{Gerv}, the
dependence on
$f_0'$ vanishes. Hence, for the case of \cref{A24} studied in this
section,   \cref{Gerv} is
consistent
with
other formulations of the LBK theorem, such as LNS
(\cref{LNS}) and Low's (\cref{Low}).

\section{LBK invariance under momenta transformation}
\label{sec:amb}
We consider here the invariance under \cref{delta} in the case where $\Delta$
arises from linear transformations of the momenta in $\cal H$. Specifically,
we prove that, at NLP, \cref{squaredn}
is invariant under the following transformation,
\begin{equation}\label{SymmTransf}
{\cal H}(p_1, \dots, p_n) \to {\cal H}(\tilde p_1, \dots, \tilde p_n)~,
\end{equation}
with
\begin{align}
\tilde p_i(k)=p_i+c_i k + \order{k^2},
\label{pk}
\end{align}
where the coefficients $c_i$ are arbitrary.
To verify the invariance, let us apply \cref{SymmTransf} to \cref{squaredn}. We
get
\begin{align}
\overline{|{\cal A}(p_1, \dots, p_n,k)|}^2
&=\sum_{ij=1}^{n}(-\eta_i\eta_jQ_iQ_j)\frac{p_i\cdot p_j}{p_i\cdot k
    \,p_j\cdot
    k}\overline{|{\cal H}(\tilde p_1,\dots,\tilde p_n)|}^2 \notag \\
& \quad +\sum_{ij=1}^{n}(-\eta_i\eta_jQ_iQ_j)\frac{p_{i\, \mu} }{p_i\cdot k }
\eta_j\left(g^{\mu\nu}-\frac{p^{\mu}_jk^{\nu}}{p_j\cdot k}\right)
\frac{d}{d
    p_j^{\nu}}\overline{|{\cal
        H}(\tilde p_1,\dots,\tilde p_n)|}^2
~,
\label{ck}
\end{align}
where partial derivatives have been replaced with total derivatives due to
the non-trivial functional dependence inside the non-radiative amplitude.
The function $\overline{|\cal H|}^2$ can be
simply
expanded in $k$ by using the functional dependence of \cref{pk}, to get
\begin{align}
\overline{|{\cal H}(\tilde p_1, \dots, \tilde p_n)|}^2
= \overline{|{\cal H}(p_1, \dots, p_n)|}^2
+ k^{\mu} \sum_i c_i \frac{\partial}{\partial p_i^{\mu}}
\overline{|{\cal H}(p_1, \dots, p_n)|}^2
+ \order{k^2}~.
\label{Htaylor}
\end{align}
To proceed further, we note that so far momentum conservation has not been
imposed. We can do so by making the dependence over the momenta $p_i$ explicit
in the soft momentum, i.e. by enforcing  $k^{\mu}\to
k^{\mu}(p_1,\dots,p_n)=\sum_i\eta_ip_i^{\mu}$. Subsequently, by
differentiating \cref{Htaylor} we get
\begin{align}
\frac{d}{d p_j^{\nu}}
\overline{|{\cal H}(\tilde p_1 ,\dots,\tilde p_n)|}^2
&=\frac{\partial}{\partial p_j^{\nu}}
\overline{|{\cal H}(p_1
    ,\dots,p_n )|}^2
+\delta_\nu^{\mu}\eta_j
\sum_i c_i \frac{\partial}{\partial
    p^{\mu}_i}\overline{|{\cal H}(p_1
    ,\dots,p_n )|}^2
+{\mathcal O}(k)~.
\label{totDer}
\end{align}
Note that since \cref{Htaylor} is expanded up to $\order{k^2}$, we have
to
truncate \cref{totDer} to $\order{k}$ because
differentiating $k$ w.r.t the momenta $p_i$ reduces the order of the expansion.
In particular, $\frac{d \order{k^2}}{d p_j^{\nu}}=\order{k}$.
This is not a problem, since the l.h.s. of \cref{totDer}
is multiplied by an expression which is suppressed w.r.t. the
other term in \cref{ck}
by one power of $k$.
Therefore, plugging \cref{Htaylor} and \cref{totDer} into \cref{ck} we get
\begin{align}
    \overline{|{\cal A}(p_1, \dots, p_n,k)|}^2
    =&\sum_{ij=1}^{n}(-\eta_i\eta_jQ_iQ_j)\frac{p_i\cdot p_j}{p_i\cdot k
        \,p_j\cdot
        k}
    \overline{|{\cal H}(p_1,\dots,p_n)|}^2 \notag \\
    & +\sum_{ij=1}^{n}(-\eta_i\eta_jQ_iQ_j)\frac{p_{i\,\mu} }{p_i\cdot k }
    \eta_j\left(g^{\mu\nu}-\frac{p^{\mu}_jk^{\nu}}{p_j\cdot k}\right)
    \frac{\partial}{\partial p_j^{\nu}}
    \overline{|{\cal H}(p_1,\dots,p_n)|}^2 \notag\\&
    + R(c_i)~,
\label{squareDelta}
\end{align}
where
\begin{align}
R(c_i)
&=\sum_{ij=1}^{n}(-\eta_i\eta_jQ_iQ_j)\frac{p_i\cdot p_j}{p_i\cdot k
    \,p_j\cdot
    k}
k^{\mu}\sum_m c_m \frac{\partial}{\partial p^{\mu}_m}
\overline{|{\cal H}(p_1,\dots,p_n)|}^2 \notag \\
& \quad +\sum_{ij=1}^{n}(-\eta_i\eta_jQ_iQ_j)\frac{p^i_{\mu} }{p_i\cdot k }
\eta_j\left(g^{\mu\nu}-\frac{p^{\mu}_jk^{\nu}}{p_j\cdot k}\right)
\eta_j \sum_m c_m \frac{\partial}{\partial p^{\nu}_m}
\overline{|{\cal
        H}(p_1,\dots,p_n)|}^2 +{\cal O}(1)
~.
\label{Delta}
\end{align}
To prove the invariance of \cref{squaredn} under \cref{SymmTransf} at NLP,
we have to show that the
remainder term
$R(c_i)$, which depends on the arbitrary coefficients $c_i$, is
${\cal O}(1)$ (i.e. NNLP). This follows
straightforwardly by first noting that, thanks to $\eta^2_j=1$, the term in the
first line of \cref{Delta} cancels with the analogous term in the second line.
The term proportional to $g^{\mu\nu}$
then vanishes since $\sum_j\eta_jQ_j=0$ by charge conservation, thus leaving
$R(c_i)={\cal O}(1)$, as
desired.

Having established that the LBK theorem in the form of \cref{squaredn} is
invariant under \cref{SymmTransf} at NLP,
the consistency of the theorem (i.e. the possibility to evaluate $\cal H$
on a set of unphysical momenta as in \cref{squaredn}) follows as a corollary.
In fact, the invariance under \cref{SymmTransf}
guarantees
that at NLP there is an infinite number of equivalent forms of the theorem,
one for each choice of the coefficients $c_i$.
In particular, the unphysical momenta of \cref{squaredn} corresponds to the
trivial transformation with
$c_i=0$. On the other hand, we can choose $c_i$ so that momentum
conservation for non-radiative amplitude is restored (as in the strategy of
Burnett and Kroll), i.e.
\begin{align}
\sum_i \eta_i c_i = -1 ~, \qquad  \sum_i \eta_i \tilde{p}_i = 0~.
\end{align}
Therefore,
the form of the theorem in \cref{squaredn} where
$\cal H$ is evaluated on a set of unphysical momenta
is equivalent, up to NNLP corrections, to the form where momentum conservation
is restored (and thus no ambiguity is present).

Note that although the argument presented in this appendix is quite
general, it fails when the invariance of the amplitude cannot be represented by
linear shifts. A simple example is given by a constant amplitude that does not
depend on the external momenta. In that case, for the consistency
of \cref{squaredn}
one has to rely on the more general argument of \cref{sec:validity}.

\section{Calculation of the modified shifts}
\label{sec:ansatz}
We show here the calculation that leads to the expression in \cref{modshifts}.
We consider the conditions (i)-(iii) of \cref{eins}, \cref{zwei} and
\cref{drei}.
The most general form for $\delta p_i$ reads
\begin{align}
\delta p_i^\mu = \sum_j A_{ij}^{\mu\nu} p_{j\nu} + B_{i}^{\mu\nu} k_\nu.
\end{align}
However, it is enough for our purposes to consider
\begin{align}
\delta p_i^\mu = \sum_j A_{ij} p_{j}^\mu + B_{i} k^\mu.
\label{ansatz}
\end{align}
We can further restrict our ansatz by assuming the set
$\{p_i^\mu, k^\mu\}$ to be linearly independent. Although clearly not true in
general, this is not a problem since we are only interested in finding a
single solution. With this assumption, we can now insert \cref{ansatz} into
 \cref{eins}, \cref{zwei} and \cref{drei} to determine the
coefficients $A_{ij}$ and $B_i$.
We get
\begin{enumerate}[label=(\roman*)]
    \item \begin{align}
    \sum_{i,j} \eta_iA_{ij}
    p_{j}^\mu +
    \left(\sum_i \eta_i B_{i}+ 1\right) k^\mu = 0 ~,
    \end{align}
    which gives
    \begin{align}
    \sum_i
    \eta_iA_{ij}
    = 0, \quad \sum_i \eta_iB_{i} =  -1~.
    \label{uno}
    \end{align}
    \item \begin{align}
    \left(\sum_j \left(\delta_{ij} + A_{ij}\right) p_{j}^{\mu} + B_{i}
    k^{\mu}\right)^2 =m^2~,
    \end{align}
    which gives
    \begin{align}
    \sum_{j,k} \left(2\delta_{ij}A_{ik}+A_{ij}A_{ik}\right)(p_{j}\cdot p_{k})
    +2\sum_j \left(\delta_{ij}B_{i}+A_{ij}B_{i}\right)(p_{j}\cdot k)=0~.
    \label{due}
    \end{align}
    \item \begin{align}
        A_{ij} &
        = -Q_i \left(\savg{{\cal S}_{\text{LP}}}\right)^{-1}
        \frac{\eta_j Q_j}{k \cdot p_j}
        + \order{k^2}~,\notag \\
        B_i &
        = Q_i \left(\savg{{\cal S}_{\text{LP}}}\right)^{-1}
        \sum_{j} \left(\frac{\eta_j Q_j}{k \cdot p_j}\right)
        \frac{p_j \cdot p_i}{p_i \cdot k}
        + \order{k}
        = - \sum_j A_{ij} \frac{p_i \cdot p_j}{p_i \cdot k} + \order{k}~.
    \label{tre}
    \end{align}
\end{enumerate}

As we can see, the conditions given by \cref{uno}, \cref{due} and \cref{tre}
are
not
too restrictive. Thus, we still
have the freedom to select a single solution by introducing a scalar
coefficient $A$ such that we have
\begin{align}
    A_{ij} = AQ_i \frac{\eta_j Q_j}{k \cdot p_j}~.
\label{scalarA}
\end{align}
Then,
the condition $\sum_i \eta_i A_{ij} = 0$
is immediately satisfied by charge conservation.
The remaining conditions now yield
\begin{align}
    \sum_i \eta_i B_i = -1~,
\label{one}
\end{align}
\begin{align}
    2A Q_i \sum_{j} \eta_j Q_j \frac{p_{i} \cdot p_{j}}{k \cdot p_j}
    - A^2Q^2_i \savg{{\cal S}_{\text{LP}}}
    + 2B_{i} (p_{i} \cdot k)
    = 0~,
\label{two}
\end{align}
\begin{align}
    A = \frac{-1}{\savg{{\cal S}_{\text{LP}}}} + \order{k^3}
    \qquad
    B_i
    =
    \frac{Q_i}{\savg{{\cal S}_{\text{LP}}}}
    \sum_j \frac{\eta_j Q_j}{k \cdot p_j}
    \frac{p_i \cdot p_j}{p_i \cdot k}
    + \order{k}~.
\label{three}
\end{align}

We can finally determine the coefficients $A$ and $B_i$.
Specifically,
for the coefficients $B_i$ we can use \cref{two}, which yields
\begin{align}
    B_{i}
    = -A Q_i \sum_{j} \eta_j Q_j
    \frac{p_{i} \cdot p_{j}}{(p_{i} \cdot k)(p_j \cdot k)}
    + \frac{1}{2}\frac{A^2Q^2_i \savg{{\cal S}_{\text{LP}}}}{p_i \cdot k}~.
    \label{B}
\end{align}
Assuming the behaviour of $A$ given in \cref{three},
the second term in \cref{B} is $\order{k}$,
so $B_i$ have the correct limit given in \cref{three}.
To determine $A$, we can use
\cref{one}, which yields
\begin{align}
    1 &
    = -\sum_i \eta_i B_i
    = A \sum_{i,j} \eta_iQ_i \eta_jQ_j
    \frac{p_{i} \cdot p_{j}}{(p_{i} \cdot k)(p_j \cdot k)}
    - \frac{A^2 \savg{{\cal S}_{\text{LP}}}}{2}
    \sum_i \frac{\eta_i Q^2_i}{p_i \cdot k} \notag \\&
    = -A \savg{{\cal S}_{\text{LP}}}
    - \frac{A^2 \savg{{\cal S}_{\text{LP}}}}{2}
    \sum_i \frac{\eta_iQ^2_i}{p_i \cdot k},
\end{align}
which is a quadratic equation. Defining $\chi = \sum_i
\frac{\eta_iQ^2_i}{p_i\cdot k}$, we find that
\begin{align}
    A =
    \frac{1}{\chi}\left(
        -1 \pm \sqrt{1 - \frac{2\chi}{\savg{{\cal S}_{\text{LP}}}}}
    \right).
\label{A}
\end{align}
Only the $+$ solution has the correct behaviour at low $k$.
Combining thus \cref{ansatz}, \cref{scalarA}, \cref{B} and \cref{A}, we find
the expression for the modified
shifts
as given by \cref{modshifts}.

\bibliography{ref.bib}


\end{document}